\documentclass[preprint,12pt]{elsarticle}

\usepackage{amssymb}
\usepackage{amsmath}
\usepackage{symbols}
\usepackage{subfigure}

\usepackage{lineno}

\journal{Energy and AI}

\begin{document}

\begin{frontmatter}
\title{Application of Reinforcement Learning for Multigroup Energy Grid Optimization for Neutron Transport Criticality Problems}

\author[label1]{Ben Whewell}
\ead{whewell@lanl.gov}
\author[label1]{Nathan Gibson}
\author[label1]{Ajeeta Khatiwada}

\affiliation[label1]{organization={Los Alamos National Laboratory},
            addressline={P.O. Box 1663}, 
            city={Los Alamos},
            state={NM},
            postcode={87545},
            country={USA}
            }

\begin{abstract}
The optimization of energy group structures is integral to ensure the accuracy of multigroup neutron transport calculations. 
This works introduces the use of reinforcement learning (RL) with surrogate modeling to optimize the group structure for one-dimensional spherical $k$-criticality problems.
The proximal policy optimization (PPO) RL algorithm is modified to be used with energy grid structures, rewarding accurate group structures while favoring fewer energy groups.
This method starts from a high-fidelity energy grid and remove energy bounds until reaching a target energy structure. 
The RL agent identify which bounds are important for the final group structure, which prevent it being stuck in local minima without limiting the initial group structure. 
Neural network surrogate models that incorporate energy, material, and spatial information are used for evaluating energy grid structures without requiring full transport simulations. 
This alleviates the computational constraint commonly used in other group structure optimization problems in addition to accelerating the RL training process. 
Applied to Godiva and BeRP ball problems, the RL constructed group structures outperform commonly used group structures.
The RL group structure optimization method is also shown to perform similar to the hierarchical agglomeration approach but offers more flexibility.  
\end{abstract}

\begin{highlights}
\item Applied reinforcement learning to multigroup group structure optimization problem for criticality problems.
\item Implemented a neural network surrogate model that incorporated the energy grid, material, and spatial information, accelerating the reinforcement learning process. 
\item Outperformed LANL30 and LANL70 group structures when applied to uranium and beryllium-reflected plutonium test problems. 
\item Allows for more flexibility and less computational costs than previously implemented group structure optimization techniques.
\end{highlights}


\end{frontmatter}


\section{Introduction}
The neutron transport equation (NTE) models how neutron populations interact and behave within different systems \cite{Lewis:1993}. 
The solution of the neutron transport equation is essential to different nuclear engineering applications, such as the design and analysis of nuclear energy systems, radiation protection and safety, and neutron radiography.
In the NTE, the neutron flux is computed numerically through deterministic (e.g. discrete ordinates) and probabilistic (e.g. Monte Carlo) methods. 
The multigroup discretization scheme integrates the energy variable over finite ranges to create piecewise constant functions where each piecewise constant referred to as an energy group. 
The selection of the integration limits, or energy group bounds, is crucial to ensure accurate numerical solutions. 
There is a delicate balance between selecting high fidelity energy grids, which leads to longer computation times and larger memory footprints, and low fidelity grids, which place a higher importance on the location of the energy bounds. 
Oftentimes, historic and generalized energy group bounds are selected based on user familiarity and available computational resources. 
These include the LANL30 (NJOY3), LANL70 (NJOY11), and LANL618 (NJOY34) group bounds \cite{Njoy:2017}, where the trailing number identifies the number of energy groups in that grid structure.

These universal energy group structures provide satisfactory flux approximations to diverse types of problems and applications. 
The multigroup error can lead to shortcomings when applying the same energy grid to different materials as the flux spectra can vary widely.
Developing more optimal group structures is a non-trivial task that is often problem specific but has been an active field of research, especially through the use of optimization and machine learning (ML) algorithms.
Different optimization algorithms such as particle swarm optimization (PSO) \cite{Fasina:2022, Yi:2013} and hierarchical agglomeration and division \cite{Berry:2022} have been investigated. 
Likewise, frameworks \cite{Rouse:2025} have looked to identify the most advantageous algorithms for families of criticality problems. 
Previous ML group optimization work has focused on identifying and assessing optimal energy group structures with neural networks \cite{Berry:2021} and random forests \cite{Saller:2023}.

This research presents a technique that uses reinforcement learning (RL) to create optimal group structures for $k$-criticality neutron transport problems. 
The application of RL agents to this novel work allows for flexibility in starting group structures while alleviating computational costs and avoiding poor locally optimal solutions.
Given training the RL model is sample inefficient, often requiring millions of time steps to train a model, a neural network based surrogate model is also introduced. 
The surrogate model, a one-dimensional convolutional neural network (CNN), is able to classify proposed energy grid structures into optimal and non-optimal categories using an error metric that incorporates the effective multiplication factor and reaction rates. 
These CNN models avoid the need for expensive transport simulations required at each time step.
This group optimization approach is more accurate than commonly used group structures, LANL30 and LANL70, for one-dimensional spherical test problems. 
It is also less computationally expensive than previously used group optimization techniques.

The organization of the paper is as follows.
The neutron transport equation, energy grid discretization, and the problems with discretization are introduced in Section \ref{sec:nte} while Section \ref{sec:prev-work} touches on previously explored group optimization work.
Sections \ref{sec:rl} and \ref{sec:surrogate} introduce the reinforcement learning and surrogate models used in this analysis. 
Section \ref{sec:results} gives results for one-dimensional Godiva and BeRP ball problems and Section \ref{sec:conclusion} gives conclusions and areas for future research.

\section{The Neutron Transport Equation} \label{sec:nte}
The neutron transport equation (NTE) is comprised of seven dimensions; three spatial $\bx \in D \subset \bbR^3$, two angular $\bsOmega \in \bbS^2$, energy $E > 0$ and time $t$ \cite{Lewis:1993}.
This equation approximates the behavior of advecting neutrons as the flux interacts with the target material, where the angular flux is represented as $\Psi(\bx, \bsOmega, E, t)$ and the scalar flux $\Phi(\bx, E, t)$ integrates the angular flux over the angular dimension $\bsOmega$.
The isotropic continuous $k$-eigenvalue NTE, 
\begin{align} \begin{split} \label{eq:nte-continuous}
    \bsOmega \cdot \nabla \, \Psi(\bx,\bsOmega,E) &+ \sig{t}(\bx, E) \, \Psi(\bx,\bsOmega,E) = \\
    & \int_{0}^{\infty}dE' \; \sig{s}(\bx, E' \rightarrow E) \, \Phi(\bx, E') \\
    &+ \frac{\chi(\bx, E)}{\keff} \int_{0}^{\infty} dE' \; \nu(x, E') \, \sig{f}(\bx, E') \, \Phi(\bx, E')
\end{split} \end{align}
estimates the generational change in neutron population where the effective multiplication factor $\keff > 1$ represents a growing population (supercritical) and $\keff < 1$ represents a subcritical system.
It is a time-independent problem with the streaming and collision terms on the left hand side and the scattering and fission sources are on the right.
Cross sections $\sigma$ measure the probability that a neutron interacts with a target material where $\sig{t}$, $\sig{s}$, and $\sig{f}$ represent the total, scattering, and fission interactions, respectively.
$\nu$ is the average number of neutrons created from a fission event and $\chi$ is the probability of the resulting energy where $\int_{0}^{\infty} \chi (\bx, E) \, dE = 1$.

The angular discretization uses the discrete ordinates (S$_N$) \cite{Chandrasekhar:1960} method which solves for the angular flux in specific directions by using quadrature sets to estimate the integral over the angle.
Assuming $\bsOmega_n$ and $w_n$ are discrete angles and weights for a quadrature rule over the sphere where $n \in \cN:= \{1, \ldots, N \}$, for any integrable function $u$ defined point-wise everywhere on $\bbS^2$,
\begin{align}
    \frac{1}{4 \pi} \int_{\bbS^2} d \bsOmega \, u(\bsOmega) \approx \sum_{n=1}^N w_n u(\bsOmega_n). 
\end{align}
The angular and spatial discretization remain constant throughout this work to directly compare the RL energy discretization method with other group structures and algorithms.

\subsection{Multigroup Neutron Transport}
\begin{figure}[!ht]
    \centering
    \includegraphics[width=0.75\linewidth]{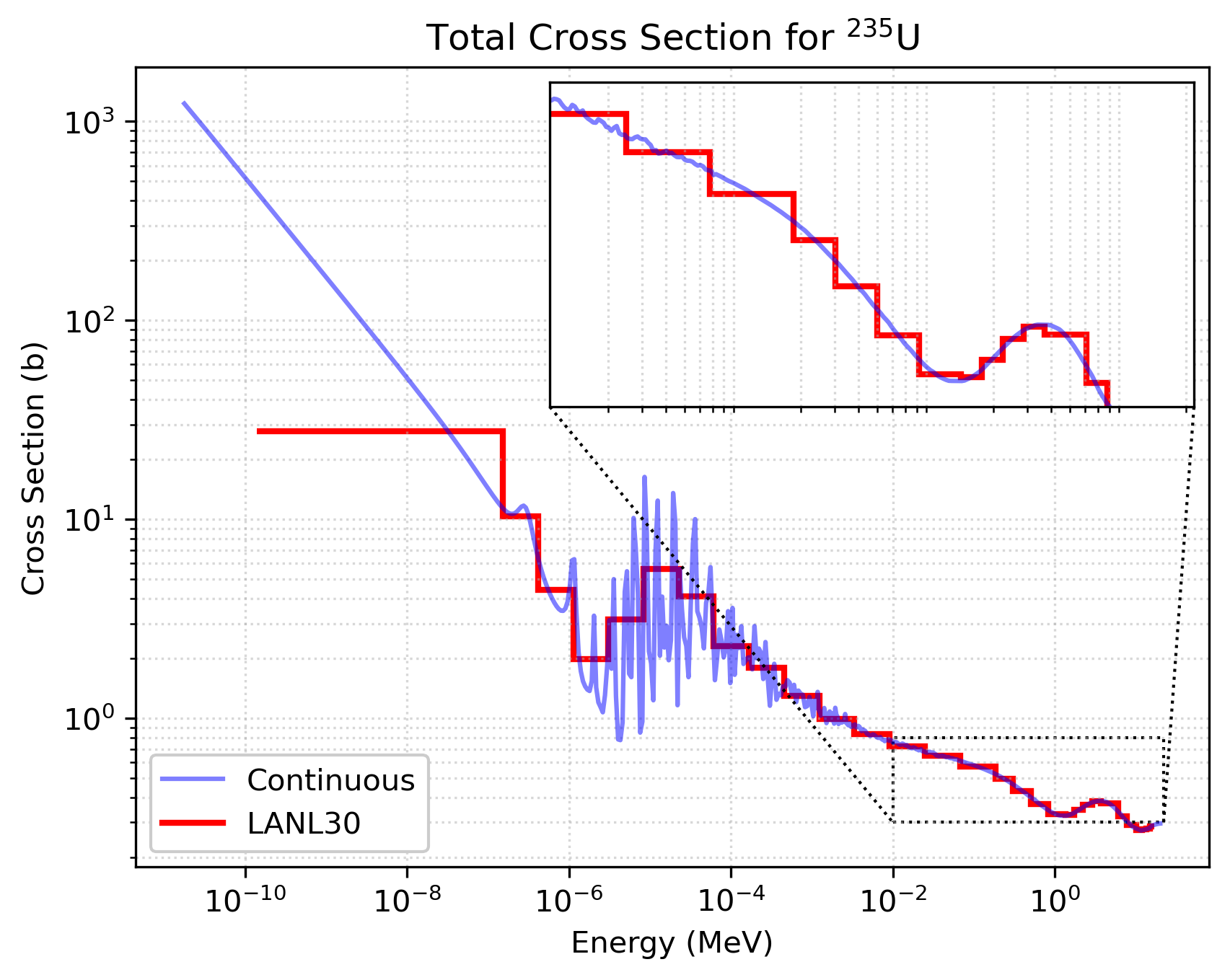}
    \caption{Comparison of the multigroup and continuous total cross section for uranium-235. The sharp peaks in the resonance region make it difficult to discretize through piece-wise constants like the LANL30 energy group structure.}
    \label{fig:xs-multigroup}
\end{figure}

The cross sections in Eq.~\eqref{eq:nte-continuous} are energy dependent and continuous, as shown in Fig.~\ref{fig:xs-multigroup}. 
Utilizing neutron cross sections in deterministic transport codes requires the discretization of the cross sections into piece-wise energy ``bins,'' referred to as multigroup cross sections.
For energy group $g \in \cG := \{1, \ldots, G\}$, the integration bounds are $E_{g-1}$ and $E_{g}$ where $E_{\rm{min}} = E_{0} < E_{1} < \cdots E_{g} < E_{g+1} < \cdots < E_{G} = E_{\rm{max}}$, which although is not standard, follows the NJOY notation \cite{MacFarlane:1987}. 
The energy grid is defined as the set $\{E_{g} \mid 0 \leq g \leq G \}$.
An approximate weighted average is also used in the discretization process shown as, 
\begin{align} \label{eq:mg-total-xs}
    \sig[g]{t}(\bx) \approx \frac{\displaystyle\int_{E_{g-1}}^{E_{g}} dE \, \sig{t} (\bx, E) \Phi (\bx, E)}{\displaystyle\int_{E_{g-1}}^{E_{g}} dE \, \Phi(\bx, E)} 
\end{align}
and 
\begin{align} \label{eq:mg-scatter-xs}
    \sig[g' \rightarrow g]{s}(\bx) \approx \frac{\displaystyle\int_{E_{g-1}}^{E_{g}}  \displaystyle\int_{E_{g'-1}}^{E_{g'}} dE \, dE' \, \sig{s} (\bx, E' \rightarrow E) \Phi (\bx, E')}{\displaystyle\int_{E_{g'-1}}^{E_{g'}} dE' \, \Phi(\bx, E')}
\end{align}
for the total and scattering multigroup cross sections.
Since the scalar flux $\Phi(\bx, E)$ is not known a priori, the weighted average must be user specified by the user through nuclear data processing software such as NJOY \cite{Njoy:2017}.

Combining the discrete ordinates method and multigroup method, the angular and scalar neutron fluxes are discretized in energy and angle as, 
\begin{align}
    \psi_{n,g}(\bx) \approx \int_{E_{g-1}}^{E_{g}} dE \, \Psi(\bx, \bsOmega_n, E) \quand  \phi_g(\bx) = \sum_{n=1}^{N} w_{n} \psi_{n,g}(\bx),
\end{align} 
for each energy group $g$ and angle $n$ where $\psi$ and $\phi$ are the discrete representations of the continuous fluxes $\Psi$ and $\Phi$.
The continuous NTE from Eq.~\eqref{eq:nte-continuous} can be updated as
\begin{align} \label{eq:nte-multigroup}
    \bsOmega \cdot \nabla \psi_{n, g}(\bx) + \sig[g]{t} \psi_{n, g}(\bx) = \sum_{g'=1}^{G} \sig[g' \rightarrow g]{s} \phi_{g'}(\bx) + \frac{\chi_{g}}{\keff} \sum_{g'=1}^{G} \nu_{g'} \sig[g' \rightarrow g]{f} \phi_{g'}(\bx)
\end{align}
for the isotropic $k$-eigenvalue transport problem with no boundary sources. 

\subsection{The Energy Grid Formulation Problem}
Neutron cross sections as functions of energy are difficult to discretize into piece-wise constants, as shown by the resonance behavior in Fig.~\ref{fig:xs-multigroup}.
The selection of these energy bound locations, $E_{1}, E_{2}, \cdots E_{G - 1}, E_{G}$, for cross section discretization is not trivial as choosing suboptimal bounds can affect the accuracy of the neutron transport calculations.
This is compounded by the fact that the optimal or near-optimal group bounds are problem specific, dependent on material compositions, geometry, orientation, and even spatial discretization resolutions. 
It also does not account for the errors within the generation of the cross sections, such as experimental errors, nuclear data uncertainties, resonance behavior, and the weighting factor selection.

It would be ideal to indiscriminately add an excessive amount of energy bounds, but this affects both the computational time and memory footprint. 
In general, more energy bounds will lead to more accurate approximations, but this is not strictly monotonic due to cancellation of errors as exemplified in Fig.~\ref{fig:godiva-ep-301}.
Fewer groups will alleviate memory and computational time but place more importance on selecting the energy group bounds.
Identifying the balance between accuracy and computational costs is the crux of group structure optimization.

There are commonly used energy grids for multigroup neutron transport simulations, such as LANL30 and LANL618 \cite{Njoy:2017} to prevent generating new grids for every simulation.
These grid structures can provide sufficient resolution for a wide variety of problems. 
Refining these grids to minimize the error between continuous and multigroup simulations for problem specific cases remains an active area of research.

\section{Previous Group Optimization Work} \label{sec:prev-work}
Different algorithms such as particle swarm optimization (PSO) \cite{Fasina:2022} and hierarchical agglomeration and division \cite{Berry:2022} have been applied to the group structure optimization problem. 
The PSO approach used different proposed group structures for a set of particles. 
After initialization, the particles are allowed to traverse the feature space, evaluate new group structures, and determine the best location for the entire swarm.
The evaluation process uses full transport simulations for each particle to compare a cost function against a reference model, in this case, LANL618.
It is an iterative process that was able to create 70 group energy grids that outperformed the LANL250 group structure while using a low number of particles and steps.

The hierarchical agglomeration (HA) algorithm \cite{Berry:2022} starts from a high fidelity energy grid, $G_1$. 
One energy bound is removed and a transport simulation is executed to evaluate the new $G_1 - 1$ grid. 
This is repeated for every energy bound, resulting in $G_1 + 1$ simulations in the first step. 
An error metrics compare these proposed group structures against a reference solution. 
The group structure is updated ($G_{2}$) by selecting the group structure in the $G_1 + 1$ set with the lowest error, where $G_2 = G_1 - 1$.
This process is repeated until the target number of energy groups $G_{N}$ is reached. 
The hierarchical division algorithm works in the opposite direction however, one energy bound is added at each step, resulting in $G_1 < G_N$. 
Both hierarchical agglomeration and division are greedy algorithms that are able to optimize group structures for a variety of problems.

Particle swarm optimization and the hierarchical methods are effective but have pitfalls. 
They both suffer from computational costs as full transport simulations are required for evaluation steps, which can become overwhelming when using many particles and steps (PSO) or starting from a high fidelity grid (HA). 
These methods are also at risk of converging to local minima (HA) or converging poorly (PSO).

\section{Reinforcement Learning} \label{sec:rl}
Reinforcement learning (RL) is a subset of machine learning that maps observations to actions to maximize a reward \cite{Sutton:2020}.
RL differs from supervised and unsupervised learning by learning directly from its environment instead of learning from previously collected data. 
The RL model (agent) takes in information about the current environment (observation, $s_t$) and performs an action ($a_t$) resulting in a reward ($r_t$) for time step $t$.
An episode is the sequence of time steps from the initial state $s_{0}$ until the terminating state $s_{T}$. 
The set of actions taken for a given observation is called the policy $\pi_{\theta}(a_t|s_t)$ where $\theta$ is the policy parameters.
The objective is to maximize the total reward for the episode, called the return. 

\subsection{Proximal Policy Optimization}
Policy optimization is one of several categories of RL algorithms.
In policy optimization, the policy gradient is estimated to optimize the parameters through gradient ascent \cite{Schulman:2017}.
Proximal Policy Optimization (PPO) is an on-policy algorithm that introduces stability by limiting the change from an old policy $\pi_{\theta_{\text{old}}}$ to a new policy $\pi_{\theta}$.
This is accomplished by using a clipped objective function
\begin{align} \label{eq:ppo-surrogate}
    L^{\text{CLIP}}(\theta) = \bbE \left[\min\left(\frac{\pi_{\theta}(a_{t}|s_{t})}{\pi_{\theta_{\text{old}}}(a_{t}|s_{t})} \, \hat{A}_{t}, \, \text{clip} \left( \frac{\pi_{\theta}(a_{t}|s_{t})}{\pi_{\theta_{\text{old}}}(a_{t}|s_{t})}, 1 - \epsilon, 1 + \epsilon \right) \hat{A}_{t} \right)\right]
\end{align}
where $\hat{A}_{t}$ is the advantage function estimator which approximates the advantage of one action over others.
$\epsilon$ is the clip range where $\epsilon \in [0, 1]$ and commonly set to $\epsilon = 0.2$. 
This algorithm was chosen for being more stable than other RL algorithms like REINFORCE. 
It has also been shown to perform well on large discrete and continuous action spaces \cite{Schulman:2017}.
The Stable-Baselines3 \cite{stable-baselines3} implementation of the PPO-Clip algorithm was used to train the RL agents.

\subsection{Energy Group Optimization Environment} \label{ssec:env}
The energy group optimization problem can be modified to fit the framework of the PPO algorithm through modifying the observation and action spaces. 
The observation space is a binary vector of length 619 ($G_\true + 1$) to represent an energy bound at that location, meaning that the LANL618 group structure is represented by a vector of all ones. 
Restricting the observation space to these group bounds means that every proposed energy group structure is a subset of LANL618. 
This is a limitation of the current design of the RL problem and it is something that could be addressed in future work. 
It should be noted however, that other LANL group structures (LANL30, LANL70) are subsets of the LANL618 energy grid \cite{Njoy:2017}.
For multiple material problems, a second vector of each material's length in centimeters is included in the observation space.

The action space for the RL agent will also be a vector of size 619, where selecting one of the actions will either add or remove the energy bound at that location, depending on the current state. 
The action space and observation space action to energy group bound is a one-to-one mapping.
Drawing inspiration from hierarchical agglomeration, the RL model will start from a high number of energy groups and remove one bound at each step until it reaches the target number of energy groups. 
To limit the agent to only remove energy bounds, action masking \cite{Huang:2020}, which sets the probability of selecting invalid actions to zero, is utilized.
This allows for future work to train and include RL agents that only add energy bounds, analogous to hierarchical division. 
In addition, expanding the action space to allow for energy bound perturbations to not be limited to generating subsets of the LANL618 group structure will also be investigated.

The starting state $s_{0}$ is initialized by selecting a random number of groups, $G_{\max} \in [200, 617]$, and sampling the starting state distribution.
A PDF was created from the LANL618 group structure and eight fixed subsets for the starting state distribution.  
This was chosen because of the limitation of the RL model only being allowed to remove group bounds and therefore initialize the starting state with bounds more frequently seen in commonly used group structures.
At each time step, one energy group bound is removed by the RL model, the new group structure is assessed, and an appropriate reward is given. 
The episode ends once a target number of energy groups is reached, represented as $G_{\min}$, where $G_{\min}$ is specified beforehand.
The relationship between the state and energy groups is $G_t = (\sum s_t) - 1$ because the state is a binary vector of size 619.

\subsection{Reward Function}
The formulation of the reward function is a crucial part of RL, as improperly ``shaping'' the reward will not allow for the RL model to learn what the user expects it to learn.
For the group structure optimization problem, the RL model needs to favor grid structures with fewer energy groups and optimal energy grids.
This can be shown as
\begin{align} \label{eq:reward-total}
    r(s_t) = \left[ w_\class, \, w_\group \right] \, \left[ r_\class(s_{t}), \, r_\group(s_t) \right]^T
\end{align}
where the proposed grid performance reward is $r_\class$, the number of energy groups reward is $r_\group$, and $w_\class$ and $w_\group$ are their associated scalar weights.
The weights allow for normalizing the individual rewards between [-1, 1] as
\begin{subequations}
\begin{align} 
    \label{eq:r-class-1} r_\class(s_{t}) &= 2 \left( \frac{y_\class(s_t) - 1}{N - 1} \right) - 1 \\[0.5em]
    \label{eq:r-group-1} r_\group(s_t) &= -2 \left(\frac{G - G_{\min}}{G_{\max} - G_{\min}}\right) + 1
\end{align}
\end{subequations}
where $y_\class(s_t)$ is the output of the discretized error category associated with state $s_{t}$, $N$ is the number of error classifications, $G_{\max}$ is the starting number of energy groups, 
\begin{align}
    G = \left( \sum s_t \right) - 1 \text{ , } \qquand G_{\max} = \left( \sum s_{0} \right) - 1
\end{align}
assuming that $s_{t}$ is a binary vector of the energy group bounds.
The specifics of evaluating an energy grid is discussed in Section \ref{sec:surrogate}.

\section{Surrogate Modeling} \label{sec:surrogate}
On-policy reinforcement learning algorithms such as PPO suffer from poor sample efficiency, requiring significant number of steps to train an RL agent \cite{Gao:2024}.
At each step in the training process, a full transport simulation must be executed to determine the performance of the proposed energy grid structure and appropriately reward the action. 
Assuming that each simulation takes on average six seconds to run and using five million time steps when training the RL model, the total simulation time would be over 8300 hours to train. 

Replacing the neutron transport simulations with a surrogate model allows for training the RL model in a more feasible manner. 
A $N = 10$ multiclass classification neural network is proposed, which takes the binary vector of length 619 ($G_{\true} + 1$) as the input and a rank classification as the output. 
For problems with multiple materials, the total cross section and material thickness are included in the surrogate model.

\subsection{Surrogate Model Architecture} \label{ssec:architect}
Two types of neural networks are used to construct the classification surrogate model, convolutional neural networks (CNNs) and Long Short-Term Memory (LSTM) networks. 
One-dimensional CNNs are used to preserve local structure found in the binary energy group bound inputs \cite{Li:2022}.
Fully connected layers join the output of the convolutional layers with the classification output. 

Multimodal neural networks are used for spatially nonhomogeneous problems, with the CNN with its binary energy group bounds input representing one modality.
A second modality utilizes LSTMs, which are often used with sequential data and can handle long-term dependencies \cite{Hochreiter:1997}. 
The LSTM encodes the total cross section and material thicknesses to allow for better generalization with different problem layouts. 
The material thickness is appended to the total cross section vector for the $G_{\true} = 618$ of each material, resulting in an input size of 619. 
Each vector of material property is viewed as sequential input where the material closest to the one-dimensional sphere center is the first material.
Fully connected layers combine the energy group bound and material property modalities, referred to as the mixture model.

\subsection{Error Metric Formulation}
The $k$-effective value is important when determining the accuracy of a neutron transport simulation, however agreement in the $k$-effective value alone does not necessarily imply agreement in the neutron flux spectra due to possible error cancellation.
In essence, a simulation can result in an accurate $k$-effective value but with an inaccurate flux spectra when compared to a reference solution because of error cancellation. 
The error metric $\varepsilon$ combines the total ($RR^{t}$), $\nu$-fission ($RR^{f}$), and absorption ($RR^{a}$) energy integrated reaction rates with the $k$-effective value. 
These reaction rates are represented as 
\begin{align}
    RR^{t} = \sig{t} \; \phi \text{ , } \qquad RR^{f} = \nu \sig{f} \; \phi \text{ , } \quad \text{ and } \quad RR^{a} = \sig{a} \; \phi
\end{align}
where $\phi$ is the energy integrated scalar flux.
These terms are combined to form the error metric
\begin{align} \label{eq:error-func} \begin{split}
    \varepsilon_{i} = \left( 
    \begin{aligned}
    \left|\frac{RR^{t}_{\true} - RR^{t}_{i}}{RR^{t}_{\true}}\right|^2 &+ \left|\frac{RR^{f}_{\true} - RR^{f}_{i}}{RR^{f}_{\true}}\right|^2 \\
     +& \left|\frac{RR^{a}_{\true} - RR^{a}_{i}}{RR^{a}_{\true}}\right|^2 + \left|3 \left(\frac{k_{\true} - k_{i}}{k_{\true}} \right)\right|^2 
    \end{aligned} \right)^{(1/2)}
\end{split} \end{align}
for a proposed energy grid $i$.
The reference values are from the LANL618 simulation results and the reaction rates are integrated over energy and space.

\begin{figure}[!ht]
    \centering
    \includegraphics[width=0.95\linewidth]{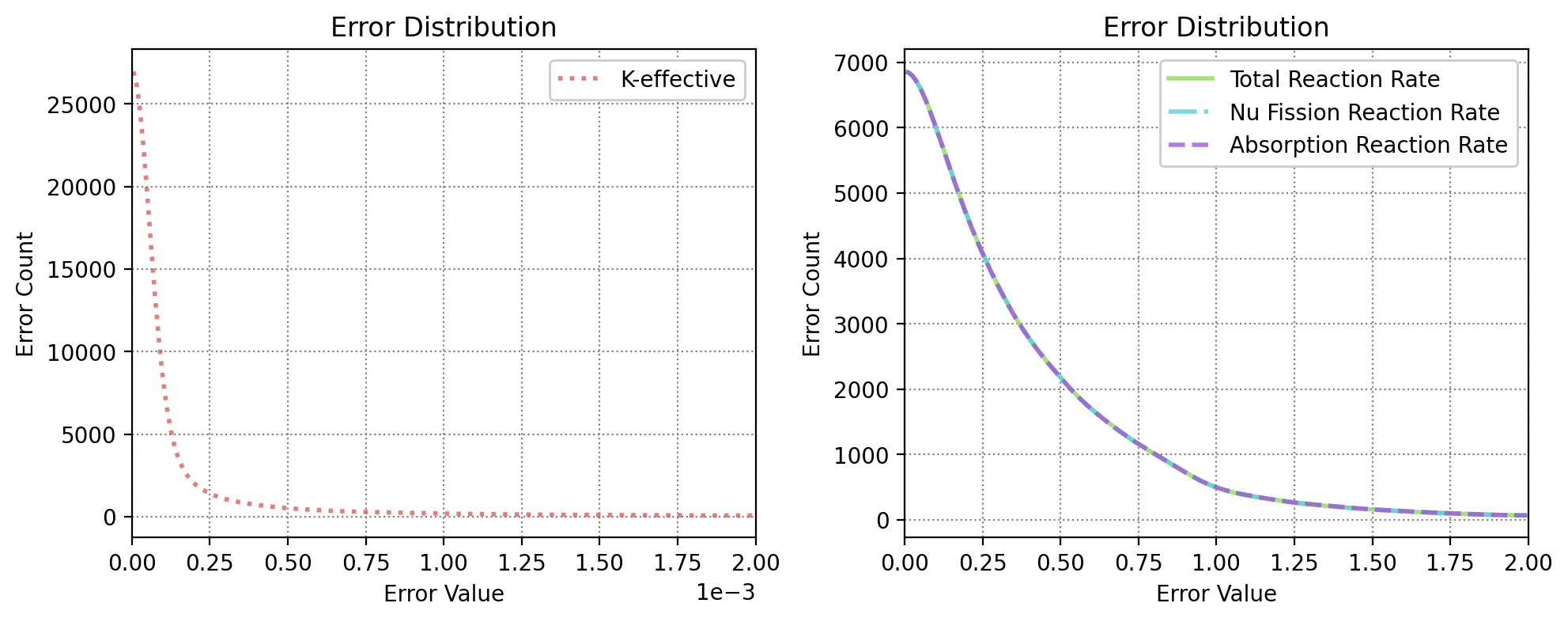}
    \caption{Error distribution for the $k$-effective, total reaction rate, $\nu$-fission reaction rate, and absorption reaction rate errors in Eq.~\eqref{eq:error-func}. Each data point comes from the Godiva training data. The $k$-effective error distribution differs in both magnitude and shape.}
    \label{fig:error-distribution-orig}
\end{figure}

\begin{figure}[!ht]
    \centering
    \includegraphics[width=0.55\linewidth]{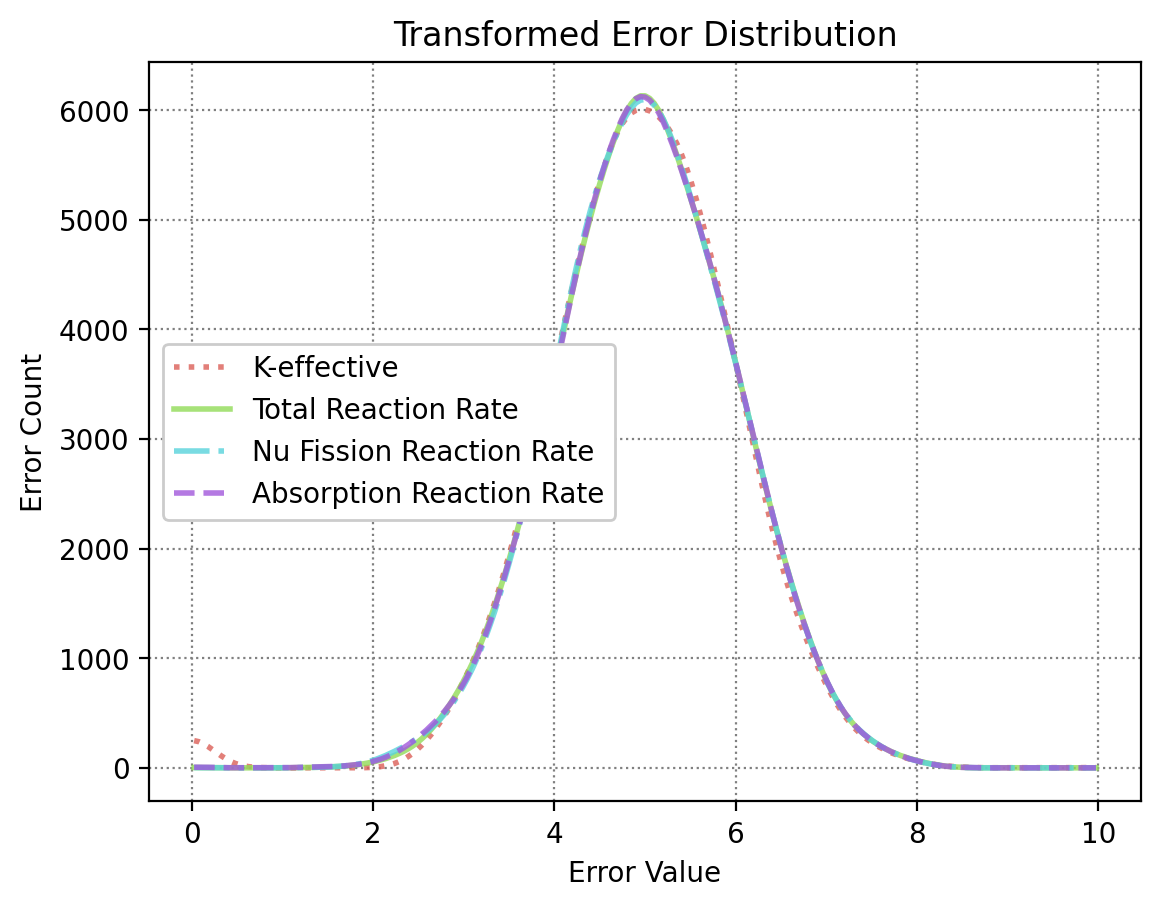}
    \caption{Transformed error distribution for the $k$-effective, total reaction rate, $\nu$-fission reaction rate, and absorption reaction rate errors in Eq.~\eqref{eq:error-func} for the Godiva training data. A quantile transformation and scaling the values between 0 and 10 were used to transform the training data in Fig.~\ref{fig:error-distribution-orig} to a normal distribution.}
    \label{fig:error-distribution-transformed}
\end{figure}

Fig.~\ref{fig:error-distribution-orig} shows that the reaction rate error distributions for the Godiva problem have similar shapes and differ significantly from the $k$-effective error distribution. 
To assign comparable weights to each error term in Eq.~\\\eqref{eq:error-func}, the relative errors are transformed into normal distributions and scaled between 0 and 10. 
The resulting transformed distributions are shown in Fig.~\ref{fig:error-distribution-transformed}. 
Additionally, the relative $k$-effective error is multiplied by 3 to give it greater significance than the combined reaction rate errors in the root-sum-square equation.

\subsection{Classification Surrogate Model}
Training data for the surrogate model is created by generating random group structures that are subsets of LANL618. 
These group structures are used in the transport code to create simulation data and its associated error.
For each proposed error, $\varepsilon_{i} \in [0, 20\sqrt{3} \approx 34.61]$ using Eq.~\eqref{eq:error-func} where $20 \sqrt{3} = [10^2 + 10^2 + 10^2 + (3 \times 10)^2]^{1/2}$, knowing the maximum error for each term is $10$. 
The distribution of error is separated into $N = 10$ classification bins, with the higher errors associated with lower classification bin numbers. 
Namely, $n = 1$ contains the least accurate (highest $\varepsilon$) energy group bound simulations while $n = 10$ contains the most accurate simulations.
The bin widths were chosen to allow for similar numbers of simulations in each bin while ensuring sufficient data in the higher classification bins for lower numbers of energy groups.

Casting this into a classification problem allows for easier training and improves the stability of the surrogate model training when coupled with the RL model.  
This method however limits the target accuracy of proposed group structures, meaning the surrogate model can only identify model accuracy based on the classification bin bounds. 
The classification bin edges consequently limit the ability of the RL model to generate the most optimal energy group structure.
This is a design tradeoff as the classification improves the surrogate robustness while introducing finite resolution in the prediction accuracy. 
The inclusion of the predicted classification probabilities also provide a more robust model output, as shown in the following section.

\subsection{Surrogate Model and Reward Function}
The reward function for the proposed group structure, shown in Eq.~\eqref{eq:r-class-1}, can be modified to fit the surrogate model. 
The surrogate model takes in the energy group bounds and material properties and outputs the probability of each error classification, $M(s_{t}) = [p_1, p_2, \ldots, p_{N}]$.
These probabilities are multiplied by the classification bin number as 
\begin{align} \label{eq:y-class}
    y_\class = \sum_{n = 1}^{N} M(s_{t})_{n} \, n
\end{align}
to account for uncertainties in the surrogate model and allows for the model output to become richer, effectively providing a probability distribution over error ranges instead of a purely discrete classification.
If the target number of energy groups is reached ($G = G_{\min}$) with the highest classification bin equating to the lowest error ($\max M(s_{t}) = N$), a success reward $r_{\mathrm{success}}$ is given. 
The updated classification reward in Eq.~\eqref{eq:r-class-1} is shown as
\begin{align} \label{eq:r-class-2}
    r_{\class}(s_{t}) = 
        \begin{cases}
        r_{\mathrm{success}} & G = G_{\min}, \; \max M(s_{t}) = N \\[0.5em]
        2 \left( \displaystyle \frac{y_\class(s_t) - 1}{N - 1} \right) - 1  & \text{Otherwise} 
        \end{cases}
\end{align}  
which is used with Eq.~\eqref{eq:reward-total}.

\section{Training and Results} \label{sec:results}
Two one-dimensional spherical geometry problems are presented, an unshielded sphere of uranium, called Godiva, and a beryllium reflected plutonium (BeRP) ball based on the International Criticality Safety Benchmark Evaluation Project (ICSBEP) benchmarks \cite{Bess:2019}. 
Data for surrogate model training is collected by selecting random group structures with energy groups $G \in [20, 618]$. 
The four error terms in Eq.~\eqref{eq:error-func} are transformed into normal distributions and the error is converted to multiclass target data with $y \in [1, 10]$.
The transformed data is used to train the surrogate model employing a 60\%/20\%/20\% train/validation/test split. 
The performance of the surrogate model is evaluated with the testing data and shown as a confusion matrix. 
Confusion matrices highlight the misclassifications of the model, where further from the diagonal is less desirable and in this case, more misclassified. 
The trained surrogate model is inserted in the reward function, described in Eqs.~\eqref{eq:reward-total}, \eqref{eq:r-group-1}, and \eqref{eq:r-class-2}, to train the RL model. 
Two RL models are trained using different target energy groups $G_{\min} = 30, 70$. 
The initial number of energy groups is varied $G_{\max} \in [200, 617]$ using the starting state distribution described in Section \ref{ssec:env}.
These trained RL models are tested with a 301 group and the LANL618 starting structures and collapsing to the target energy groups.
The RL models are verified by running the full transport simulations at each time step to represent the true error metric. 
They are compared to the LANL30 and LANL70 energy group structures as well as the hierarchical agglomeration method.
The training of the surrogate and RL models took approximately six hours each on an Apple MacBook Pro equipped with an Apple M3 Max chip (16-core CPU) and 64 GB of unified memory running Python 3.12.

\subsection{Godiva}
The Godiva surrogate model was trained on subcritical data ($k \approx 0.928$) however, it is accurate for supercritical ($k \approx 1.070$) as well by varying the radius. 
Due to this model's ability to accurately estimate proposed group structures for different material lengths, only the energy group bounds were used as the input to the surrogate model.
It was trained on 323902 different energy grid structures, with 200000 data points with random groups $G \in [20, 600]$, 99565 points with $G \in [30, 618]$, and 24337 points with $G \in [20, 70]$. 
The addition of the 24,337 data points was to allow for additional training data points in the lower energy region.
The resulting neural network used two convolutional layers with 208 and 32 filters and average pooling. 
There were three fully connected hidden layers that had 3588, 788, and 222 nodes with LeakyReLU activation and dropout layers to prevent overfitting. 
A batch size of 32, learning rate of $10^{-4}$, 50 epochs, a cross-entropy loss function, and the Adam Optimizer were used.

\begin{figure}[!ht]
    \centering
    \subfigure[]{\centering \includegraphics[width=0.495\textwidth]{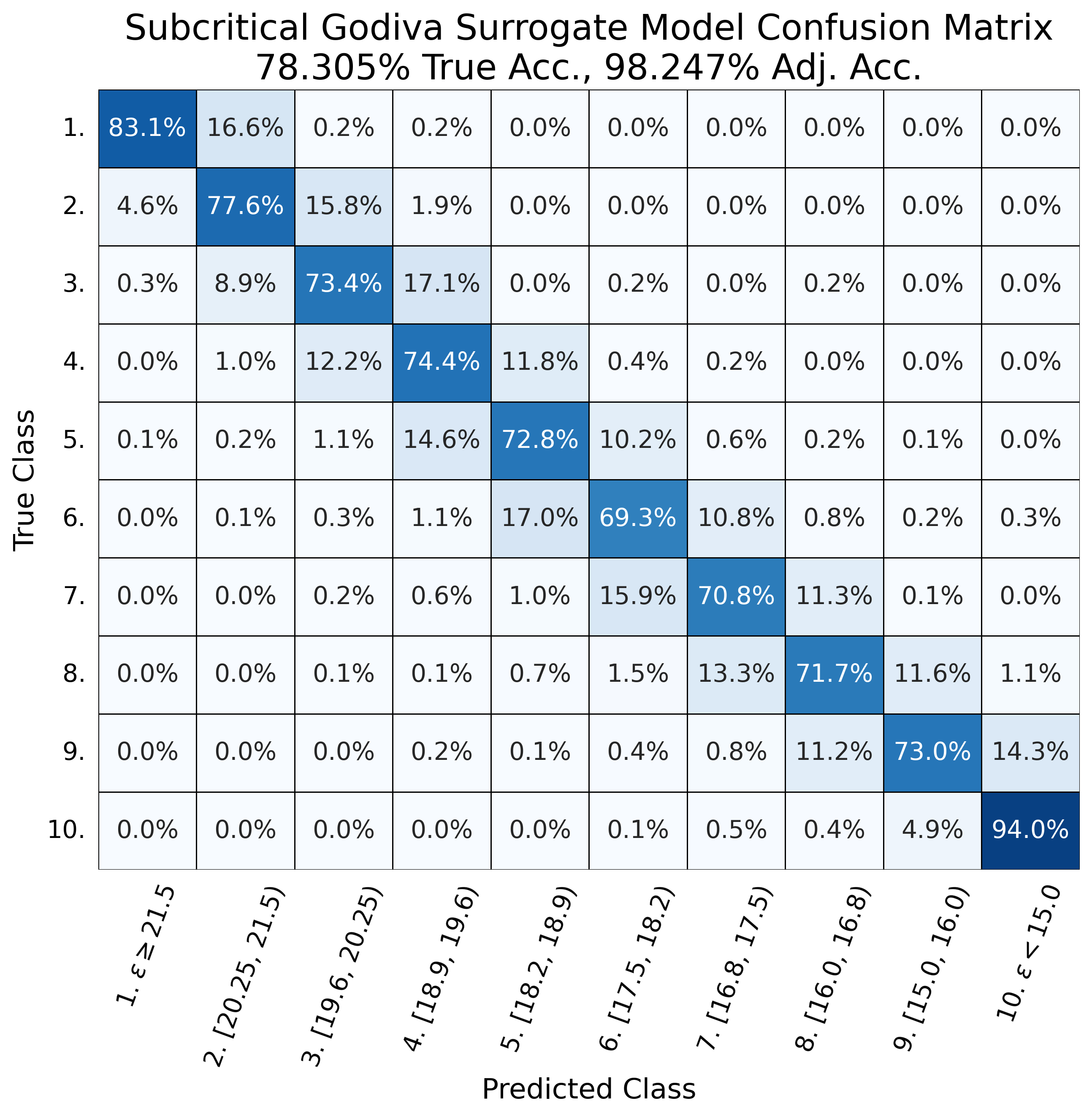}\label{fig:godiva-sub-confusion}}
    \subfigure[]{\centering \includegraphics[width=0.495\textwidth]{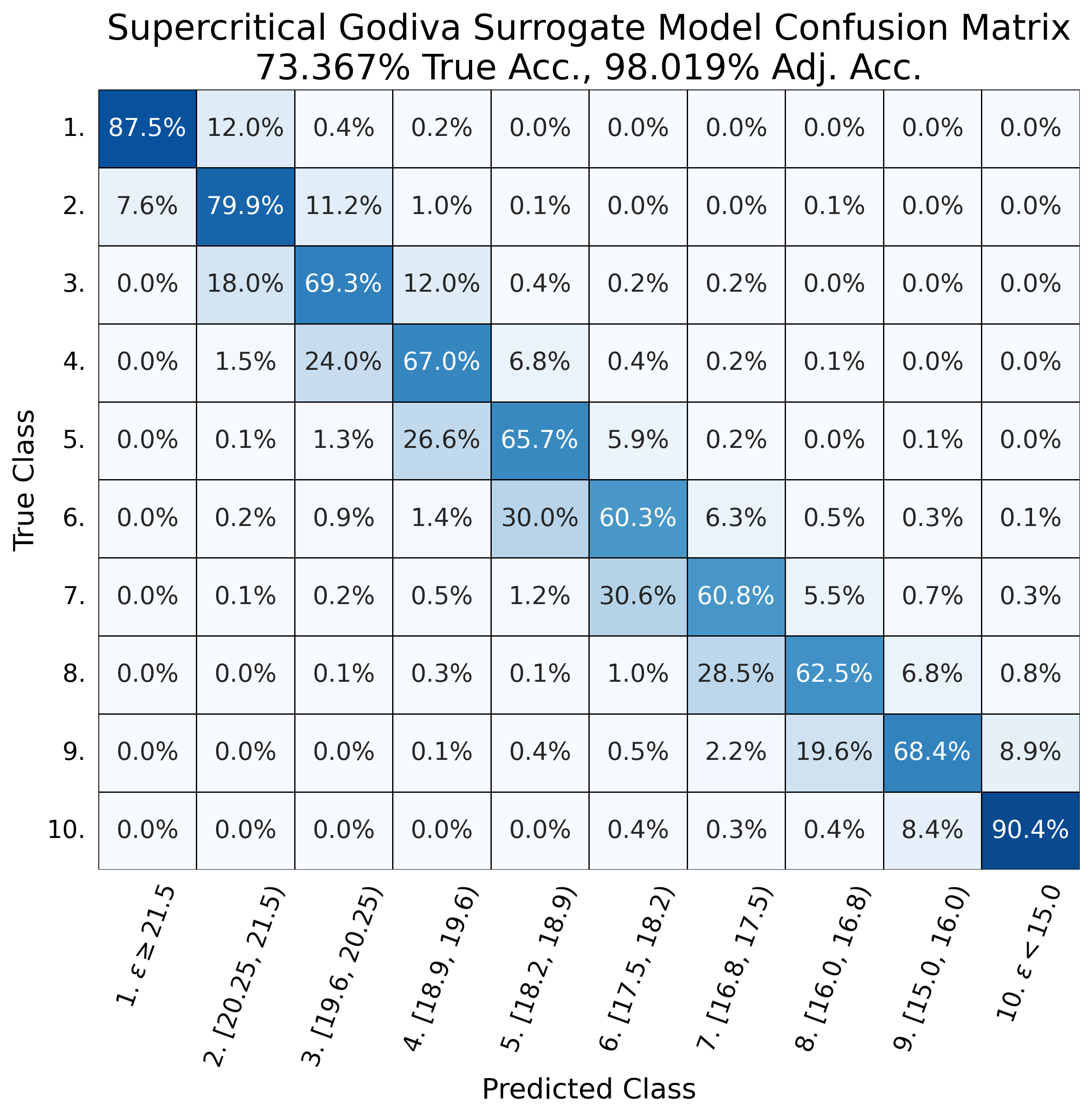}\label{fig:godiva-sup-confusion}}
    \caption{The confusion matrix for the Godiva surrogate models on test data showing predicted vs. true classifications for random group structures. The adjacent accuracy is reported due to the ordinal relationship between the classes. (a) The subcritical ($k \approx 0.928$) Godiva problem with 9984 samples $G \in [30, 618]$. (b) The supercritical ($k \approx 1.070$) Godiva problem with 9995 samples $G \in [30, 618]$.}
    \label{fig:godiva-confusions}
\end{figure}

Testing on 9984 subcritical samples $G \in [30, 618]$, the surrogate model had a true accuracy of 78.305\% and an adjacent accuracy of 98.247\%, as shown in Fig.~\ref{fig:godiva-sub-confusion}. 
The adjacent accuracy is comprised of the true accuracy and the accuracy of having a prediction within one class of the true value, represented by the tri-diagonal in the confusion matrix. 
The adjacent accuracy was included given the ordinal relationship between the classes and the formulation of the RL reward function. 
The surrogate was also tested on the supercritical Godiva problem with 9995 samples $G \in [30, 618]$, by increasing the radius of the subcritical Godiva simulation.  
For a supercritical version, the true and adjacent accuracies of 73.367\% and 98.019\%, shown in Fig.~\ref{fig:godiva-sup-confusion} demonstrate the surrogate model can generalize between different Godiva criticalities.
The performance of the surrogate model, coupled with the incorporation of ordinal relationships and predicted class probability distributions included in the reward function (Eq.~\eqref{eq:y-class}), allow the RL model to effectively optimize energy group structures beyond exact class-level agreement.

\begin{figure}[!ht]
    \centering
    \subfigure[]{\centering \includegraphics[width=0.495\textwidth]{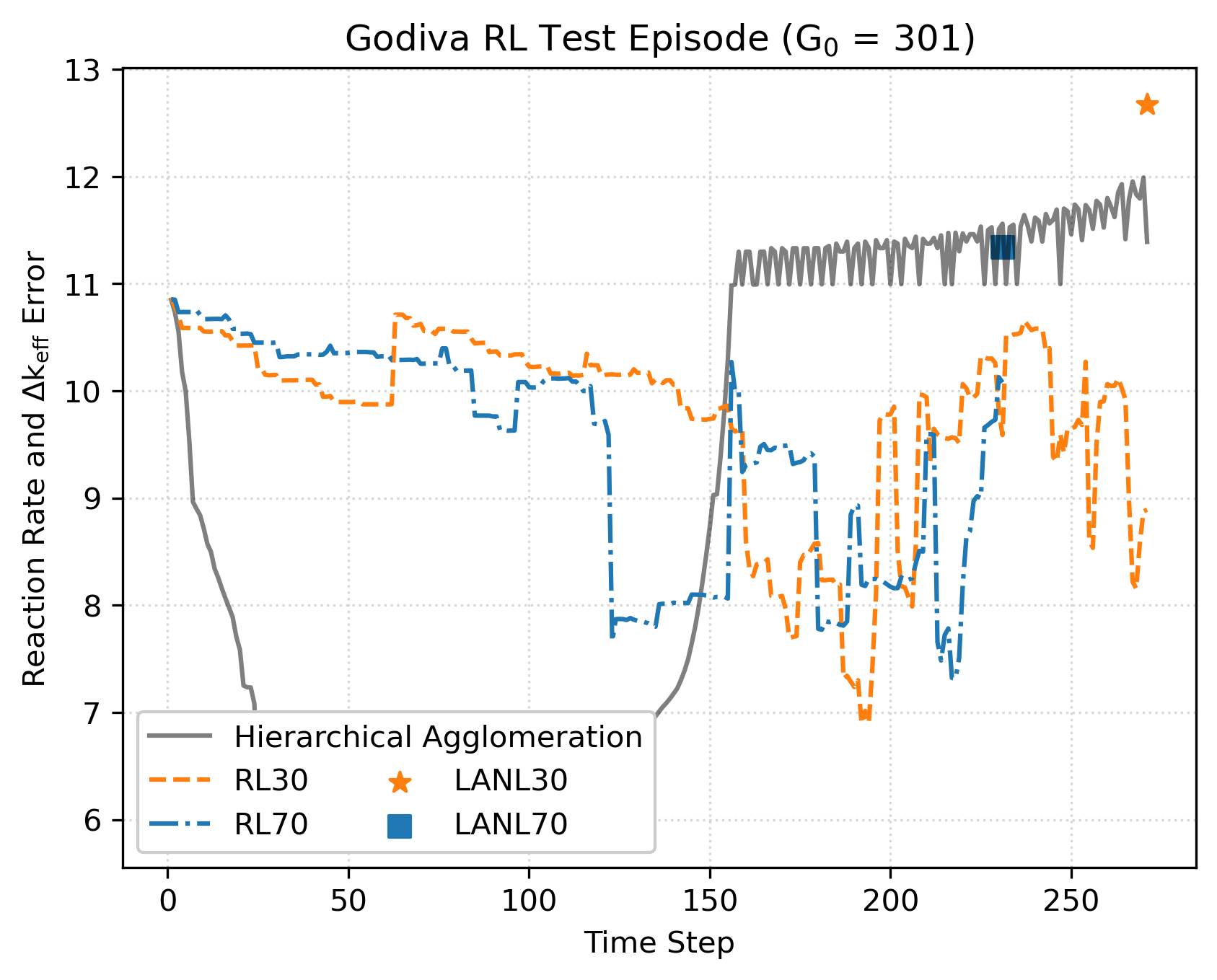}\label{fig:godiva-ep-301}}
    \subfigure[]{\centering \includegraphics[width=0.495\textwidth]{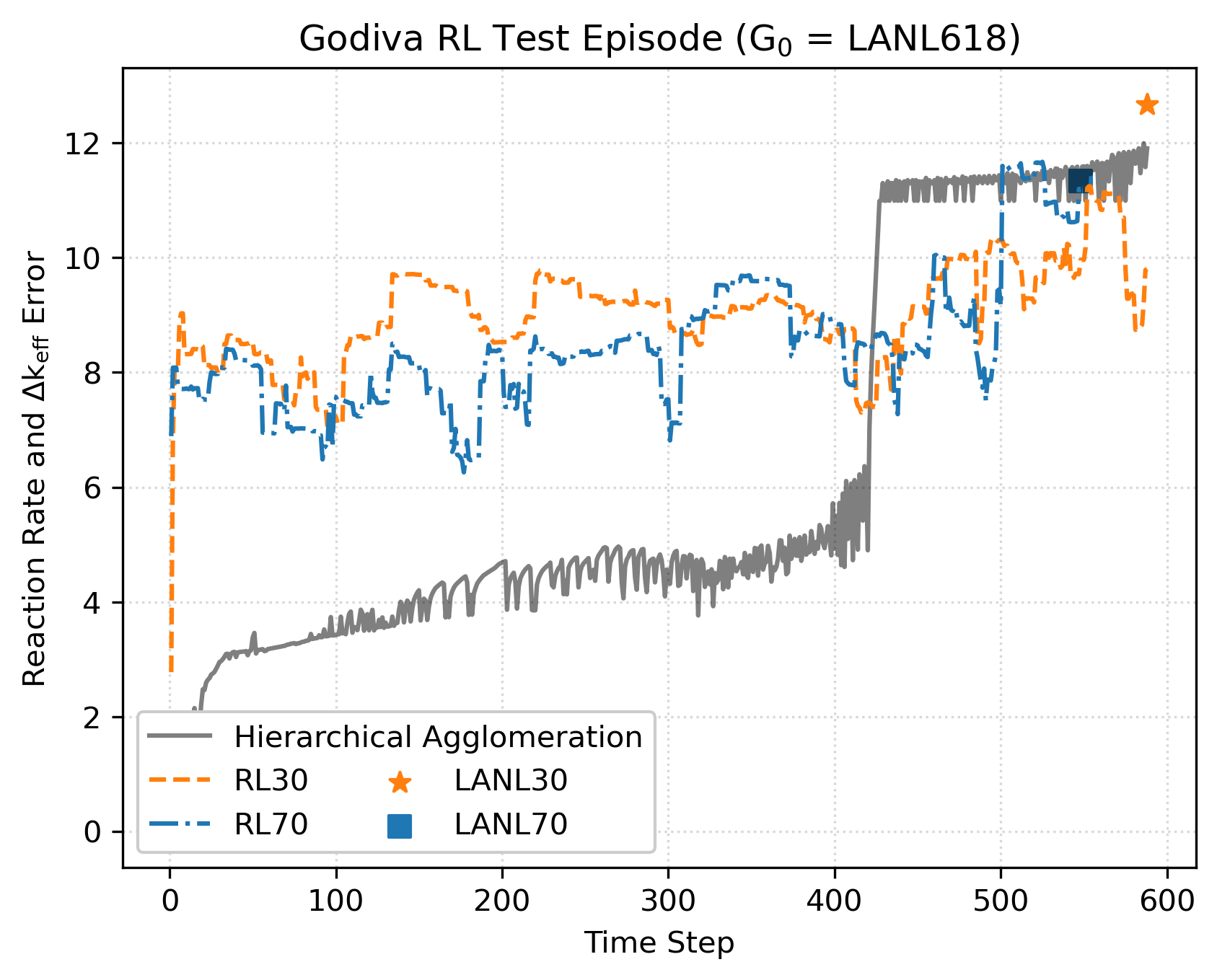}\label{fig:godiva-ep-lanl618}}
    \caption{Test episode for the Godiva RL models compared to LANL30 and LANL70. The hierarchical agglomeration group optimization structure was also used. At each time step, the true group structure was calculated and compared to the reference structure using Eq.~\eqref{eq:error-func}. (a) uses a 301 group starting structure and (b) uses the LANL618 starting structure.}
    \label{fig:godiva-ep-combined}
\end{figure}

The RL model was trained with five million time steps with the initial number of groups $G \in [200, 617]$. 
The reward weight in Eq.~\eqref{eq:reward-total} used the weights $[w_\class, w_\group] = [1, 0.001]$ and the success reward $r_{\mathrm{success}} = 400$ to allow for the final group structure to be as significant as every step in the episode. 
The PPO implementation from Stable Baselines was used with default values expect for the learning rate which was $10^{-4}$. 
The RL models were trained with target groups of $G_{\min} = 30$, $70$ and noted as RL30 and RL70, respectively. 
Test episodes used a 301 group and the LANL618 as starting structures with the results shown in Fig. \ref{fig:godiva-ep-combined}.

\begin{figure}[!ht]
    \centering
    \subfigure[]{\centering \includegraphics[width=0.495\textwidth]{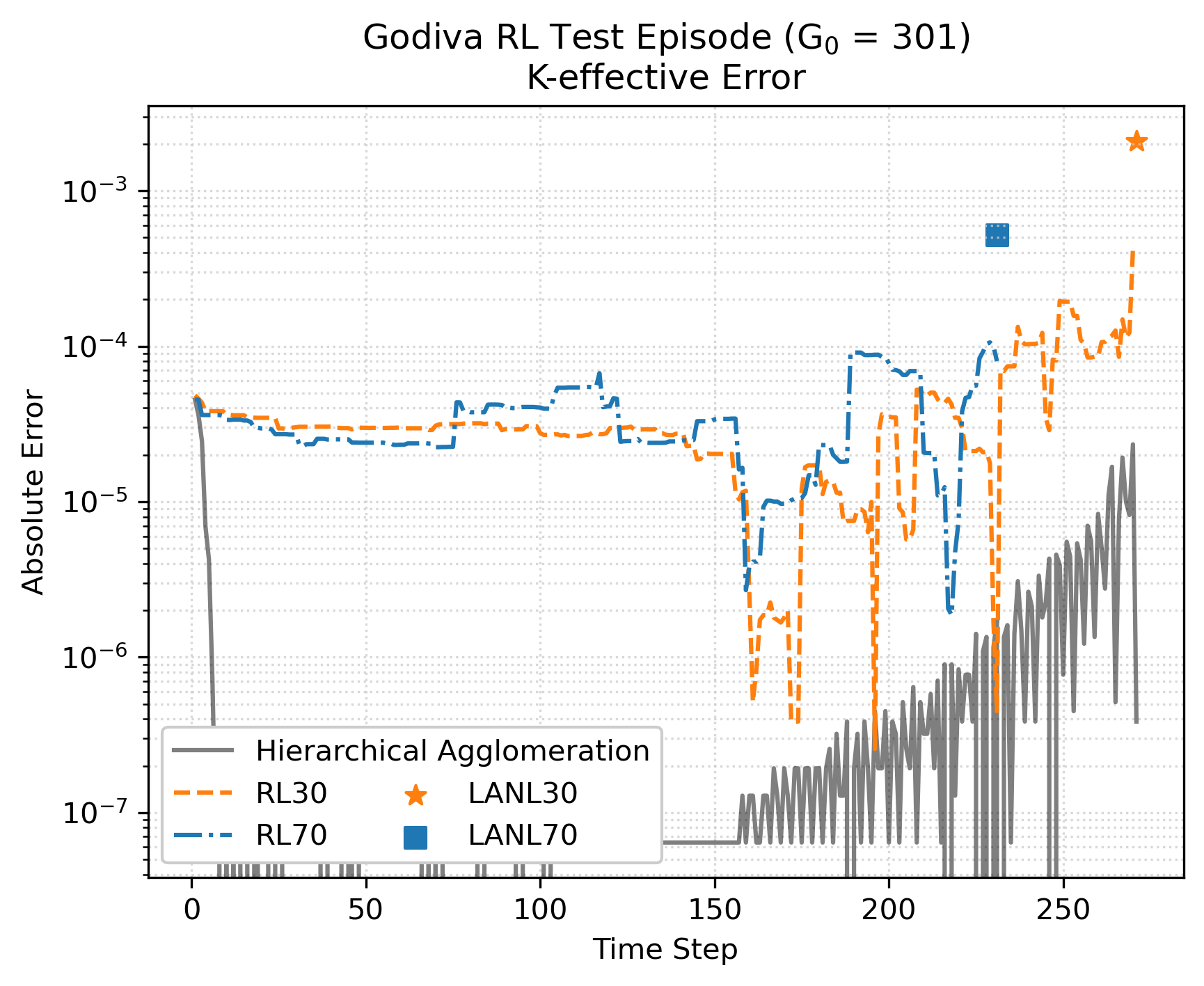}\label{fig:godiva-ep-301-tot}}
    \subfigure[]{\centering \includegraphics[width=0.495\textwidth]{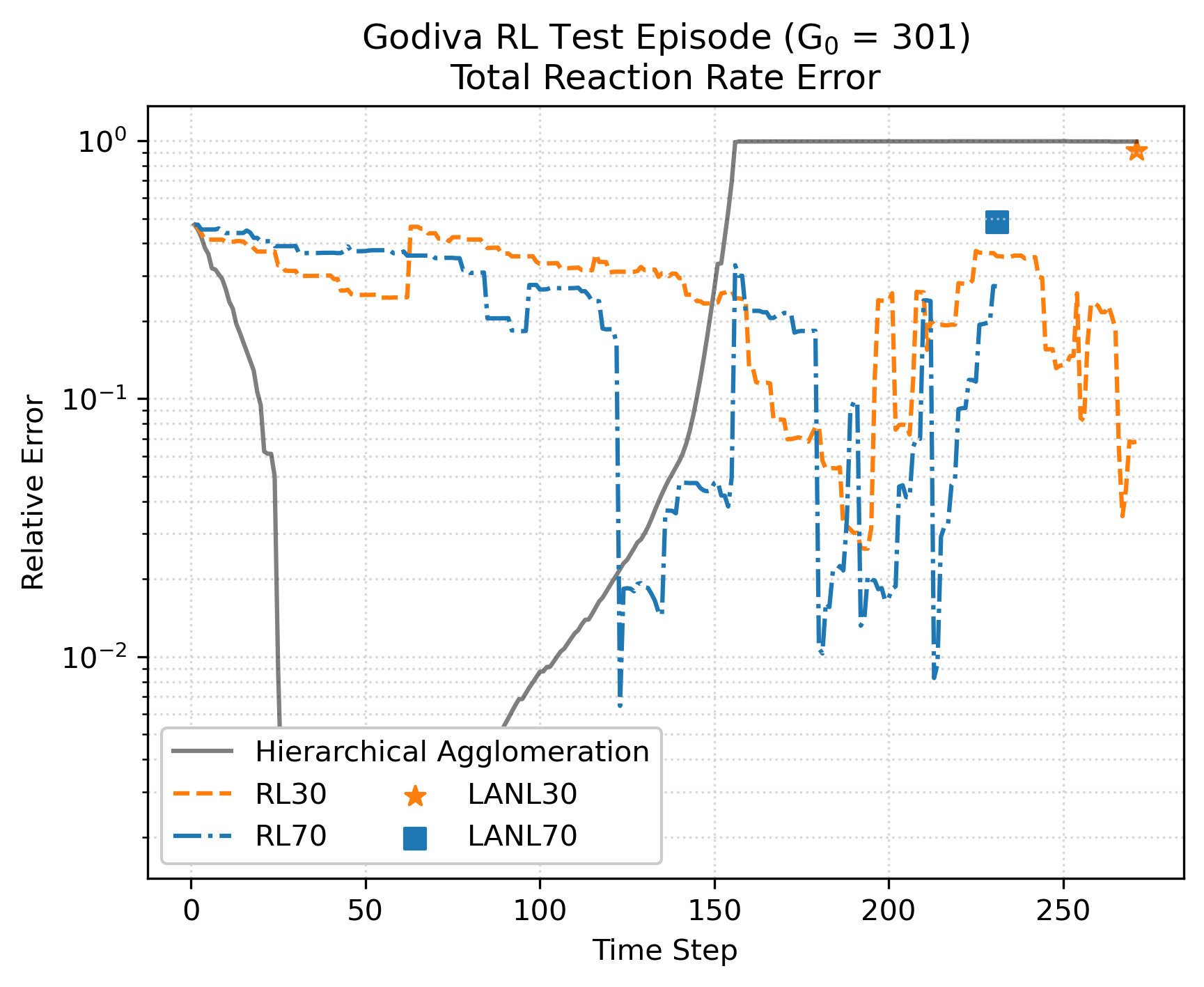}\label{fig:godiva-ep-301-keff}}
    \subfigure[]{\centering \includegraphics[width=0.495\textwidth]{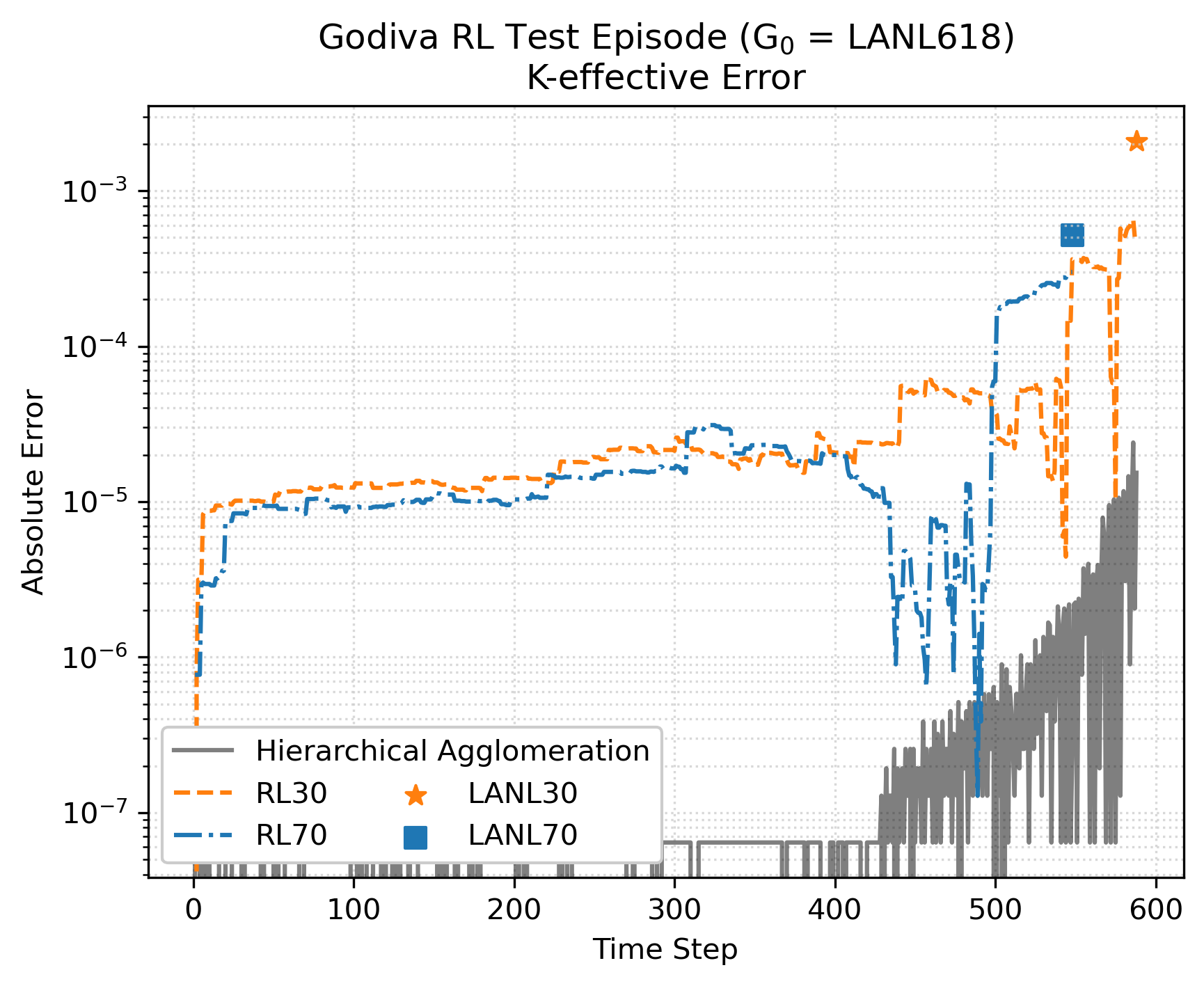}\label{fig:godiva-ep-lanl618-keff}}
    \subfigure[]{\centering \includegraphics[width=0.495\textwidth]{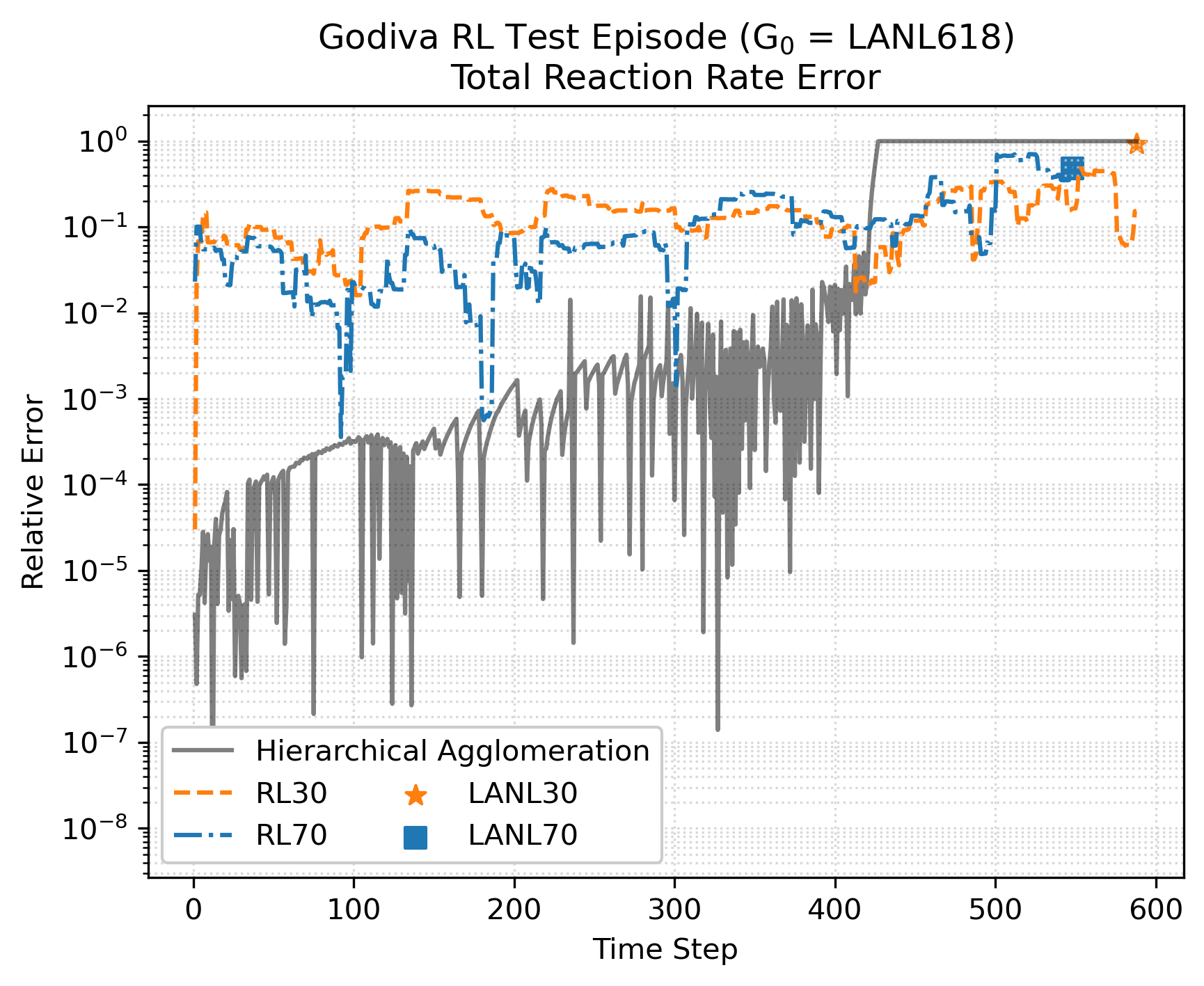}\label{fig:godiva-ep-lanl618-tot}}
    \caption{Test episodes from Fig. \ref{fig:godiva-ep-combined}. These results show the total reaction rate and $k$-effective error metrics used in Eq. \eqref{eq:error-func}. (a, c) show the absolute $k$-effective error while (b, d) show the relative total reaction rate error.}
    \label{fig:godiva-ep-separate}
\end{figure}

These test episodes show the true error using neutron simulations at each time step after removing one energy bound from the previous state.
The oscillations in both the RL30 and RL70 trajectories are caused by the limitations of the surrogate model, as every state proposed by the RL models is classified in the most accurate classification bin with a finite width.
The final step shows the RL-constructed energy grids have a more accurate error metric than their associated LANL30 and LANL70 group structures. 
The RL results were also compared with the hierarchical agglomeration (HA) method which did not use the surrogate model and instead used full transport simulations. 
HA, being a greedy algorithm, will identify the best possible move instead of identifying the best ending group structure.
This is shown with the immediate reduction in error which slowly increases with the progression of time steps. 
The oscillations are most likely cause by cancellation of errors, while the steep increase in error is caused by local minima and converging to a suboptimal solution.
It should be noted that the RL30 models in Fig.~\ref{fig:godiva-ep-301} and \ref{fig:godiva-ep-lanl618} were the same model, demonstrating the the versatility of using starting group structures. 
For comparison, every new starting group structure for the hierarchical agglomeration would require an additional simulation.

The performance of the final group structure of the RL70 model is similar to the performance of the LANL70 group structure in Fig.~\ref{fig:godiva-ep-lanl618}.
This is because the LANL70 is already in the most accurate classification bin and the RL70 model performs equally well. 

The error terms in Eq.~\eqref{eq:error-func} can also be separated into their individual terms and shown for each episode. 
The total reaction rate and $k$-effective errors are shown for the 301 group and LANL618 starting structures in Fig. \ref{fig:godiva-ep-separate}.
The $\nu$-fission and absorption reaction rates were not included as their trajectories are similar to the total reaction rate error.

The optimization objective includes multiple competing error terms and the RL models do an effective job balancing these quantities.
The HA approach appears to favor the $k$-effective term more strongly at the expense of the reaction rate preservation.
This is exemplified in the comparison of the individual $k$-effective and total reaction rate error terms  in Fig.~\ref{fig:godiva-ep-separate}. 

\begin{figure}[!ht]
    \centering
    \includegraphics[width=0.75\linewidth]{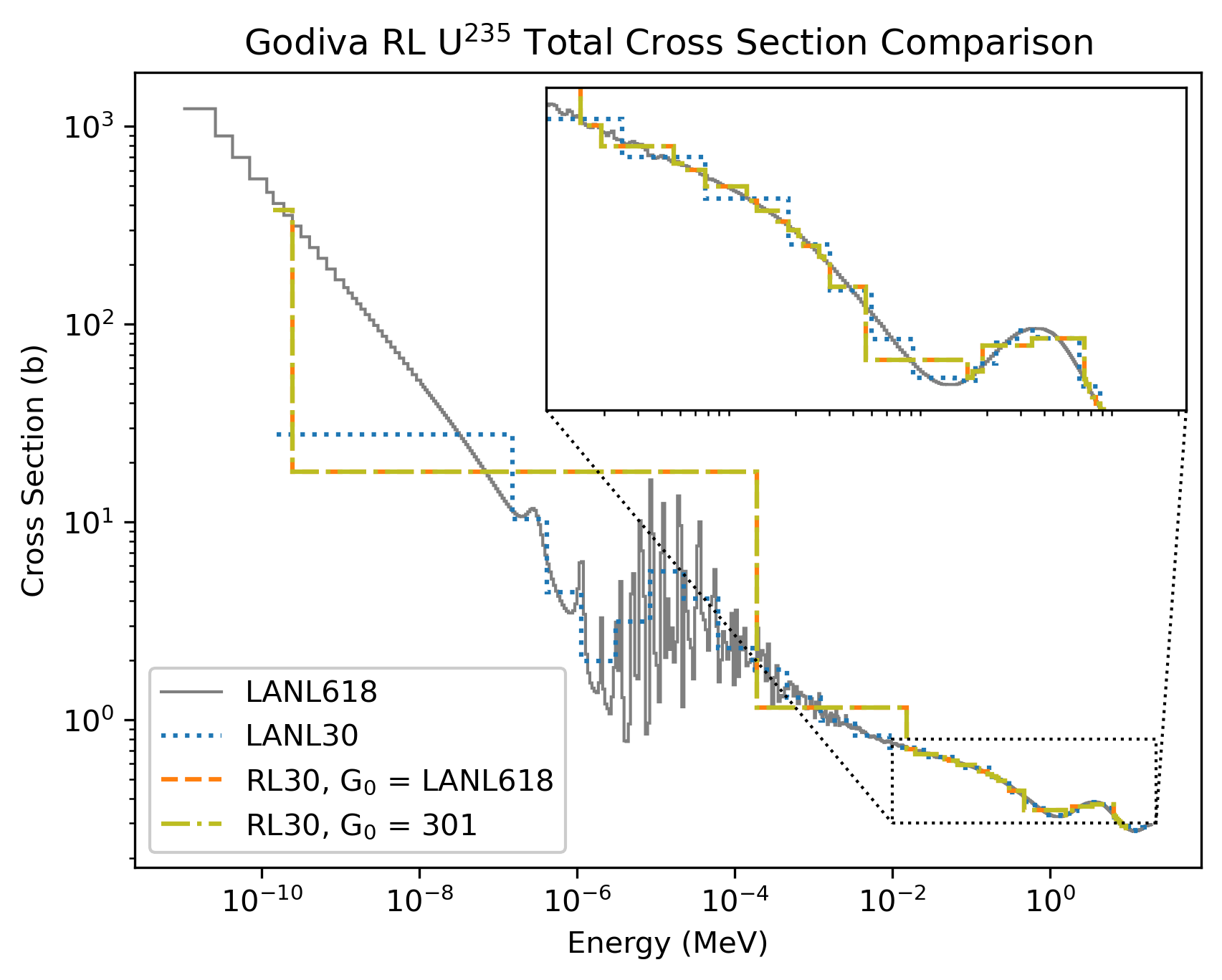}
    \caption{Comparison of the uranium total cross section for the Godiva problem using RL. The RL30 obtained from different starting group structures but have similar shapes, favoring the fast region. The LANL30 result was also compared.}
    \label{fig:godiva-xs-comparison}
\end{figure}

The final $G = 30$ group structures are also compared for the uranium total cross section in Fig.~\ref{fig:godiva-xs-comparison}.
This figure compares the RL30 results ($G_{0}$ = 301, LANL618) with the LANL618 and LANL30 group structures.
The LANL30 structure places greater resolution in the resonance region, which may be advantageous for problems dominated by thermal and epithermal neutrons.
Godiva, however, is a fast-spectrum problem, meaning most neutrons occupy the higher-energy regions.
This behavior is reflected in both RL30 group structures, where a majority of the energy group bounds are concentrated in the fast-energy region.
Fewer group bounds are placed in the thermal and resonance regions because these energy ranges are less significant for the Godiva problem \cite{Wimett:1960}.
The resulting group structures therefore demonstrate that the RL model adapts the energy discretization to the underlying neutron spectrum while improving upon commonly used reference group structures.

\subsection{BeRP Ball}
The BeRP ball is a one-dimensional sphere of plutonium surrounded by a shell of beryllium. 
Different plutonium radii were used to train the surrogate model by varying the plutonium radius between 3.5 and 4.5 cm with 0.1 cm intervals while keeping the overall length constant at 11 cm. 
This caused the problem to vary between a subcritical ($k \approx 0.922$) and a supercritical ($k \approx 1.137$) configurations.
Randomly sampled energy group structures and plutonium radii were collected resulting in 614937 data points with 286937 data points taken with random groups $G \in [20, 600]$ and the remaining 328000 data points taken with $G \in [20, 70]$.

\begin{figure}[!bh]
    \centering
    \includegraphics[width=0.75\linewidth]{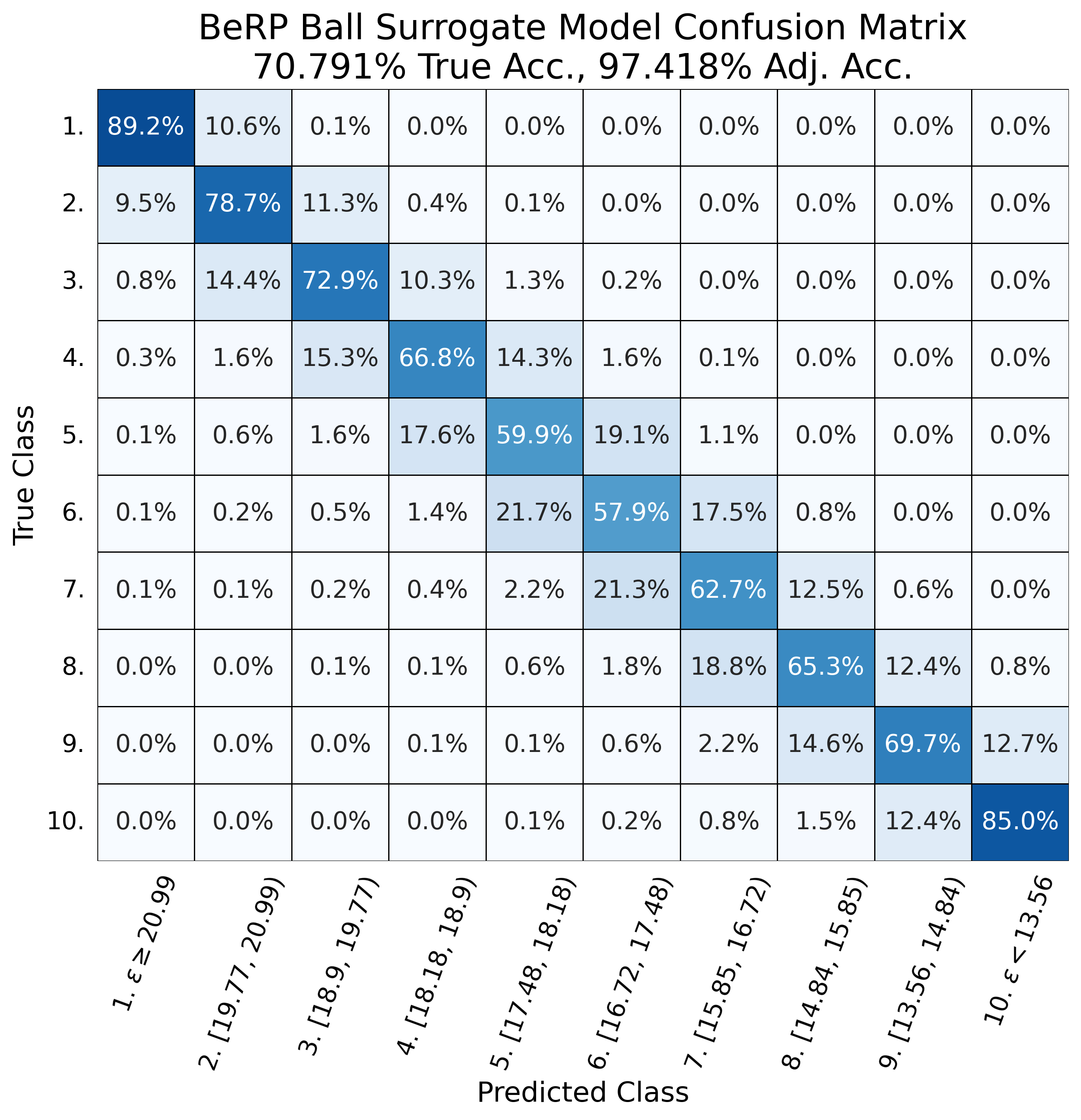}
    \caption{The confusion matrix for the BeRP ball surrogate models on test data showing predicted vs. true classifications for random group structures with $G \in [20, 600]$. The radius of plutonium was varied between 3.5 and 4.5 cm to include both subcritical ($k \approx 0.922$) and supercritical ($k \approx 1.137$) data in the test set. The adjacent accuracy is reported due to the ordinal relationship between the classes.}
    \label{fig:berp-confusion}
\end{figure}

A multimodal neural network was built to include the spatial and material information into the surrogate model in the manner discussed in Section \ref{ssec:architect}.
The energy group input used two one-dimensional CNN layers with 128 and 64 filters with max and average pooling, and one fully connected layer with 1146 nodes and LeakyReLU activation. 
The spatial and material information used an LSTM with a 3 node fully connected layer. 
The outputs from the CNN and LSTM modes were combined using a fully connected layer with one 574 node hidden layer and LeakyReLU activation.
A batch size of 150, learning rate of $5 \times 10^{-6}$, 300 epochs, cross-entropy loss, and the Adam Optimizer were used.
A learning rate scheduler and dropout layers were used to prevent overfitting.

The trained surrogate model was tested on 122988 samples with $G \in [20, 600]$ and varied the plutonium material length between $r_{Pu} \in [3.5, 4.5]$.
The surrogate model had a true accuracy of 70.791\% and an adjacent accuracy of 97.418\%, as shown in Fig.~\ref{fig:berp-confusion}. 
These results demonstrated that the surrogate model effectively captured the relative performance of proposed energy group structures within the RL reward function.
An added benefit was the inclusion of spatial and material information, which eliminated the need for new surrogate models when modifying the material layout.

The RL model was trained with ten million time steps with the initial number of groups $G \in [200, 617]$. 
The success reward $r_{\mathrm{success}} = 400$ and reward weights $[w_\class, w_\group] = [1, 0.001]$ were used. 
Default PPO values were used for the hyperparameters except for the learning rate which was $10^{-4}$. 
The RL models were trained with target groups of $G_{\min} = 30$, $70$ and noted as RL30 and RL70, respectively. 

\begin{figure}[p]
    \centering
    \subfigure[]{\centering \includegraphics[width=0.46\textwidth]{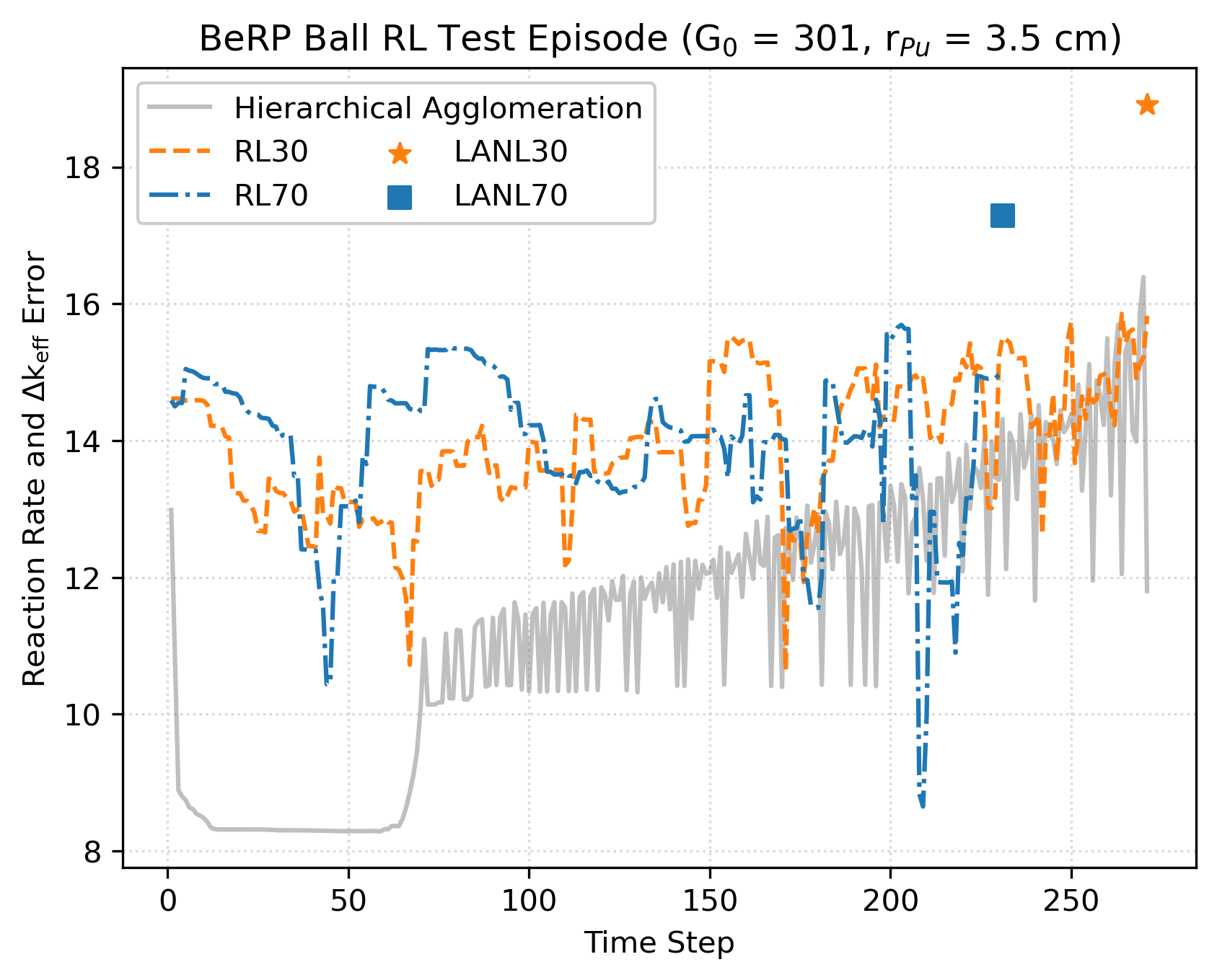}\label{fig:berp-ep-301-a}}
    \subfigure[]{\centering \includegraphics[width=0.46\textwidth]{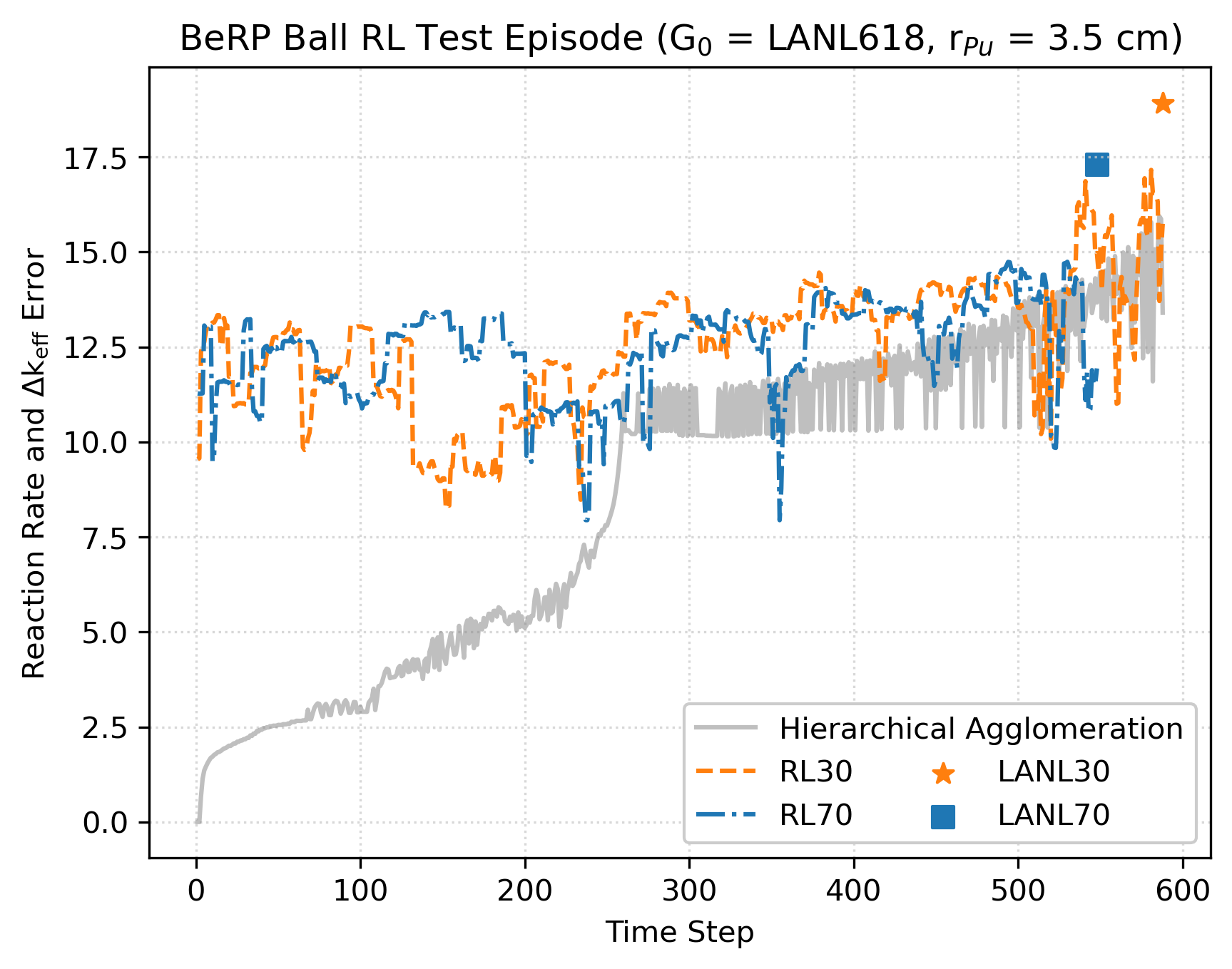}\label{fig:berp-ep-lanl618-a}}
    \subfigure[]{\centering \includegraphics[width=0.46\textwidth]{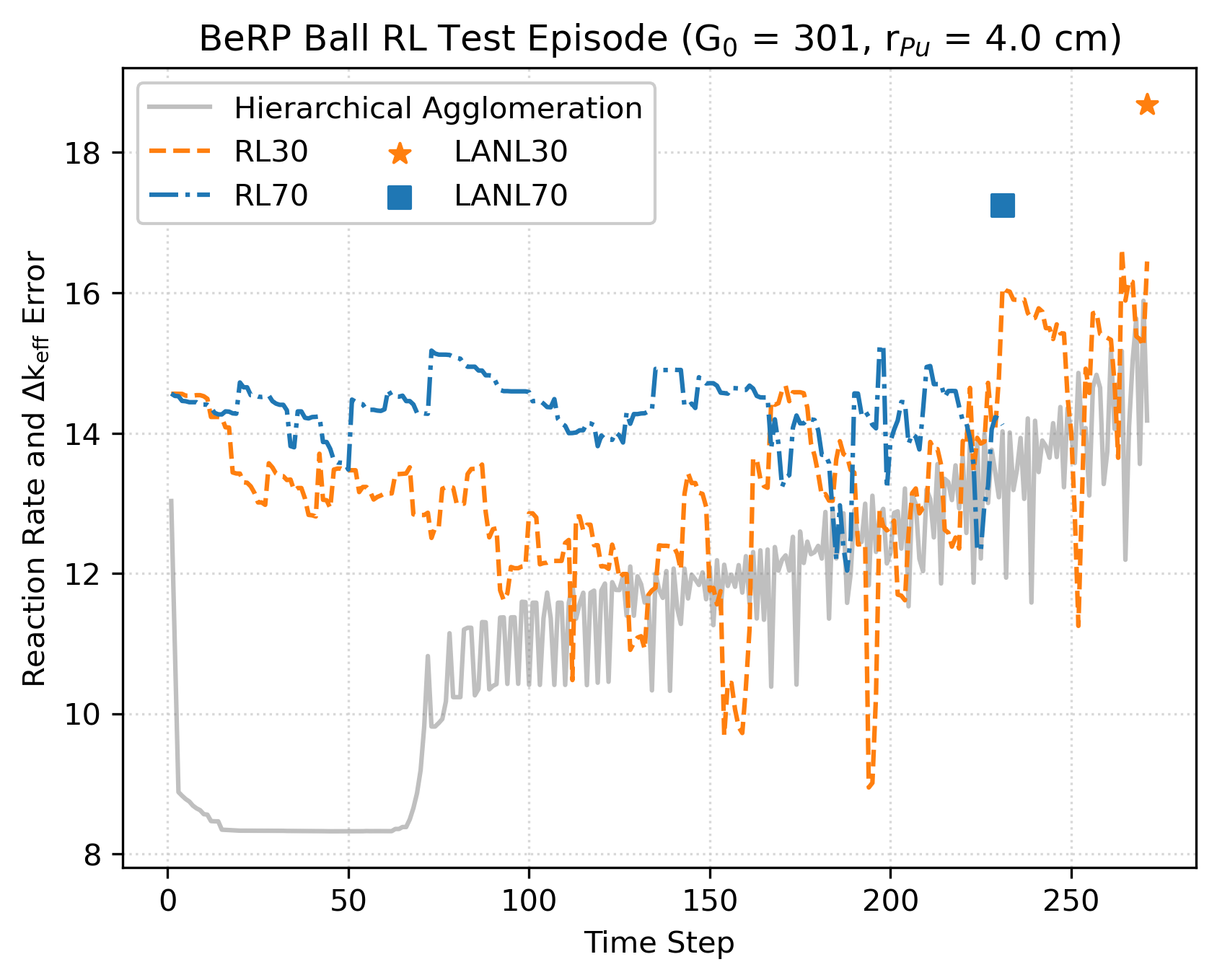}\label{fig:berp-ep-301-b}}
    \subfigure[]{\centering \includegraphics[width=0.46\textwidth]{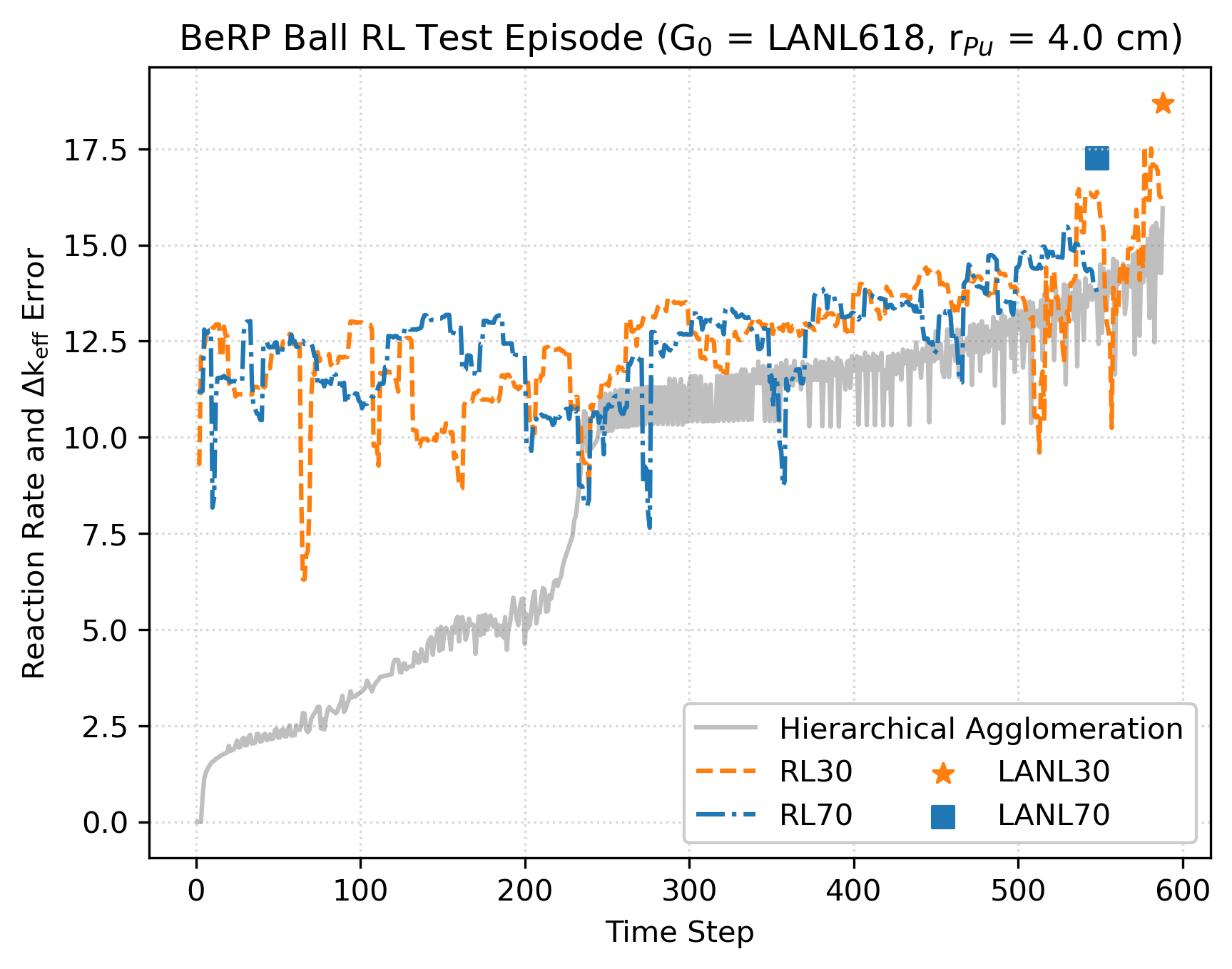}\label{fig:berp-ep-lanl618-b}}
    \subfigure[]{\centering \includegraphics[width=0.46\textwidth]{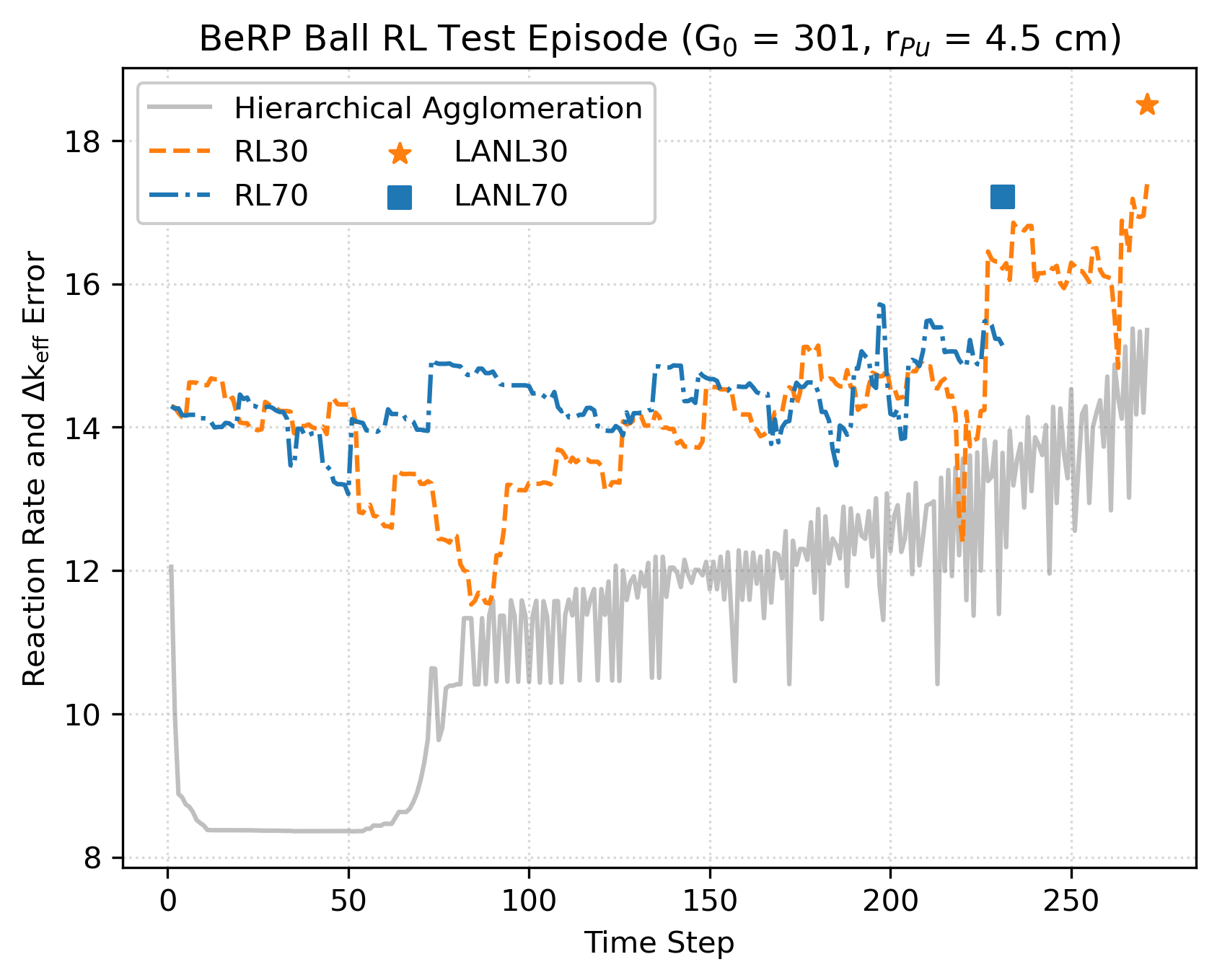}\label{fig:berp-ep-301-c}}
    \subfigure[]{\centering \includegraphics[width=0.46\textwidth]{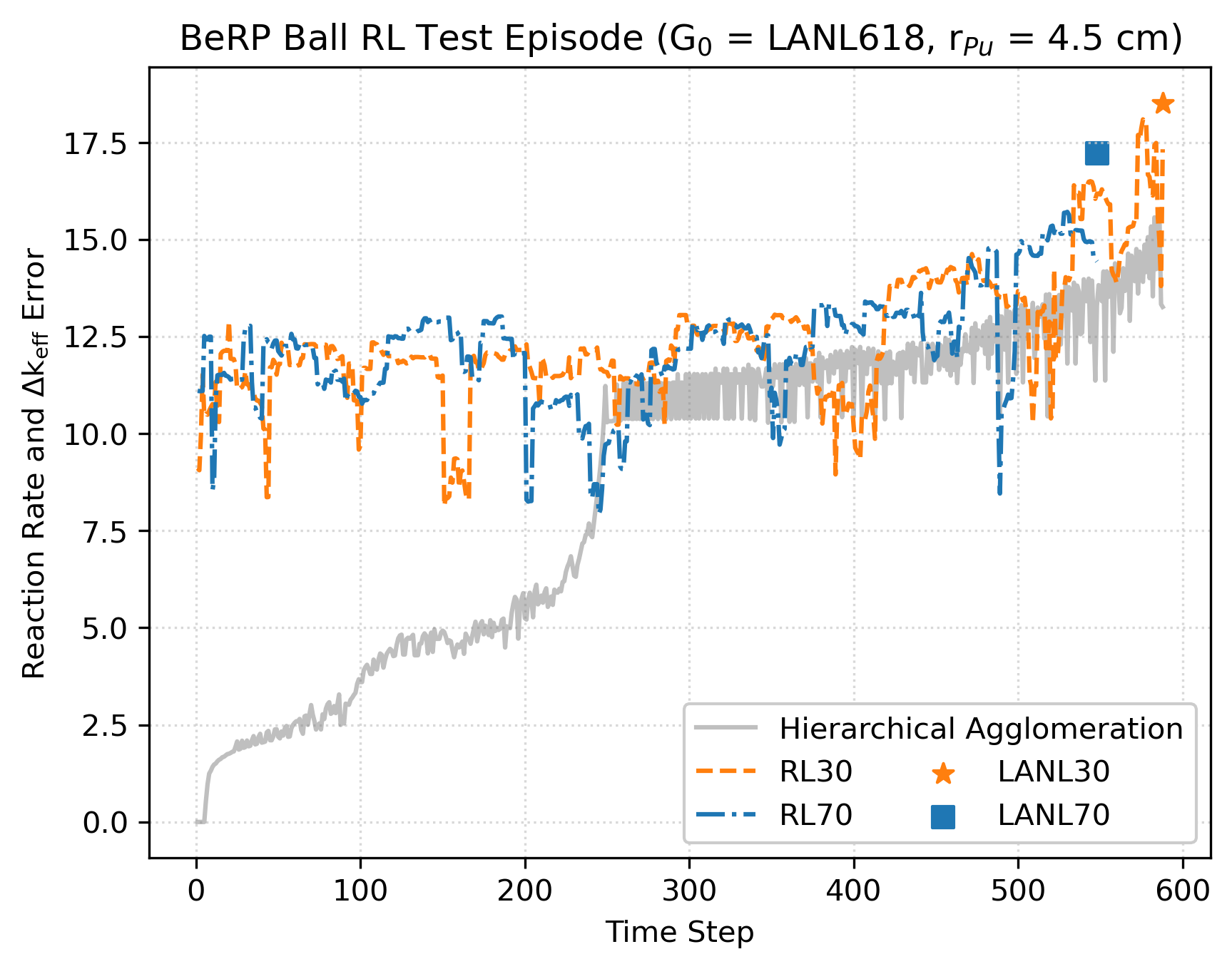}\label{fig:berp-ep-lanl618-c}}
    \caption{Test episode results for the RL BeRP ball model compared to the LANL30 and LANL70 group structures. Hierarchical agglomeration was also compared. At each time step, the true group structure error was calculated. (a, c, e) use $G_{0}$ = 301 groups while (b, d, f) use $G_{0} = $ LANL618. (a, b) use a BeRP ball with 3.5 cm of plutonium ($k \approx 0.922$), (c, d) with 4 cm ($k \approx 1.034$) and (e, f) with 4.5 cm ($k \approx 1.137$).}
    \label{fig:berp-ep-combined}
\end{figure}

The observation space was expanded to include spatial information as explained in Section \ref{ssec:env}.
To include the spatial information in the RL model, the plutonium width was sampled at the beginning of each episode with $r_{Pu} \in [3.5, 4.5]$ cm in addition to sampling the number of energy groups. 
The inclusion of the spatial information allows the RL model to generate new energy grids without training new models.
Test episodes used a 301 group and the LANL618 as starting structures and varied the plutonium widths where $r_{Pu} = 3.5$, 4, and 4.5 cm to include subcritical and supercritical examples.
These results are shown in Fig. \ref{fig:berp-ep-combined}.

One energy bound is removed at each time step in Fig.~\ref{fig:berp-ep-combined}, showing the episode history of the RL30 and RL70 models. 
The RL30 and RL70 models outperform the LANL30 and LANL70 group structures at the final timestep and in most cases, perform similarly to the HA results, such as in Fig.~\ref{fig:berp-ep-301-b}. 
The HA method requires the execution of 45225 and 191362 full transport simulations, run in semi-parallel, for the 301 group and LANL618 starting group structures, respectively. 
The HA results must be rerun when modifying the starting group structure or the material layout. 
The RL method, on the other hand, required training only two models, one with a target structure of 30 groups and one with a target structure of 70 groups. 
The RL model does not have to be retrained when modifying the starting group and material layout.
It should also be noted that the surrogate model training data for the RL method can be generated in fully parallel simulations.
Refining the size of the surrogate model classification bins or using residual RL (\cite{Johannink:2018}) will increase the performance of the RL-generated group structures when compared to other group structures and algorithms.

\begin{figure}[!ht]
    \centering
    \includegraphics[width=0.75\linewidth]{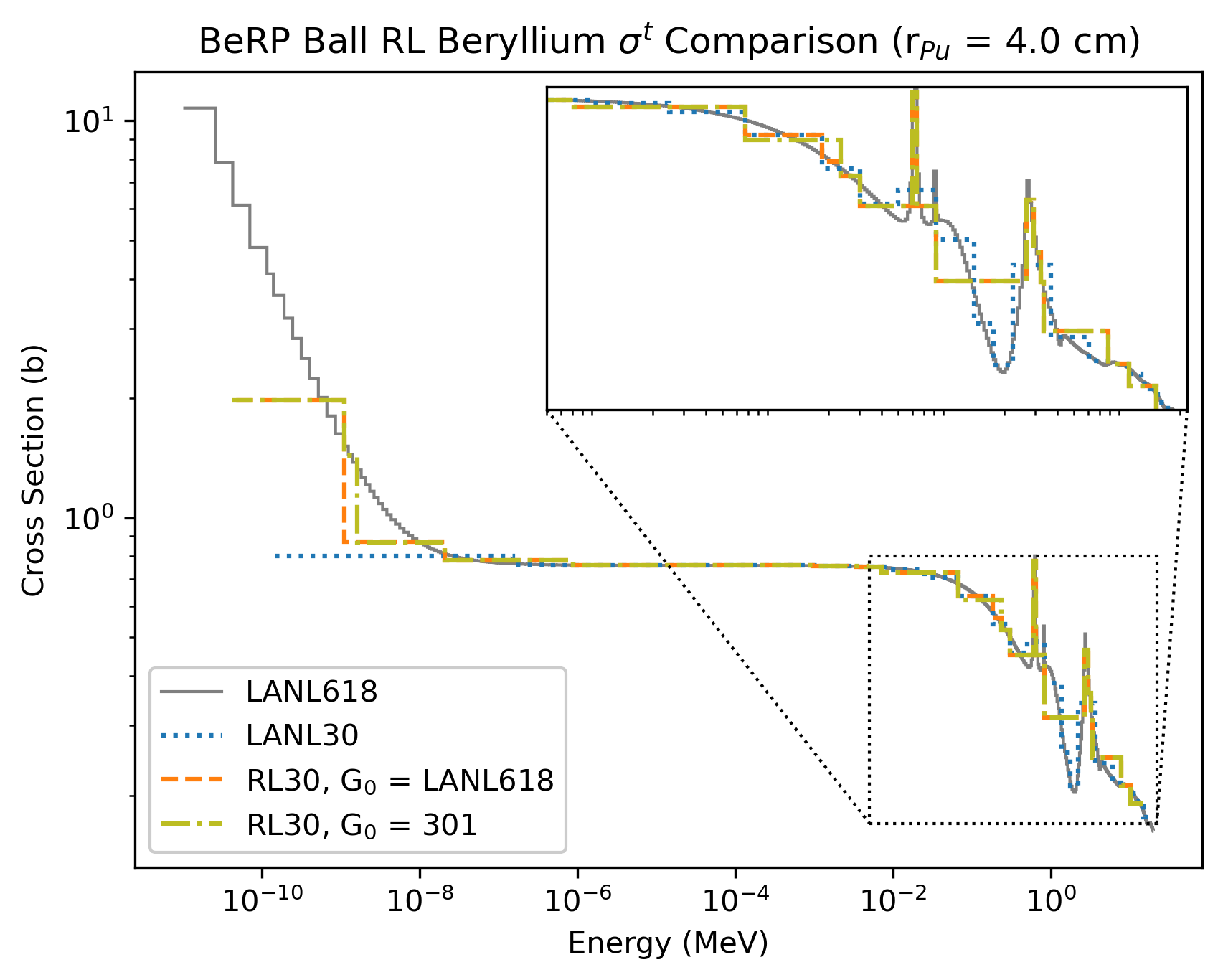}
    \caption{Comparison of the beryllium total cross section for the BeRP ball using RL for a supercritical problem ($k \approx 1.034$). The RL30 results show different starting group structures but have similar shapes, favoring the fast region and selecting some of the peaks seen in the LANL618 results. The LANL30 result was also compared.}
    \label{fig:berp-xs-comparison}
\end{figure}

The new RL30 energy group structure is compared for the total beryllium cross section, shown in Fig.~\ref{fig:berp-xs-comparison}.
While the LANL618 does not show similar resonances as the uranium cross section in Fig.~\ref{fig:godiva-xs-comparison}, the RL models continue to favor the fast regions of the energy spectra.
The RL30 model was able to capture some of the peaks present in the fast region, regardless of the starting group structure.
Focusing on the fast regions and capturing meaningful structures in the total cross section demonstrate that the RL models are learning the correct behavior.

\section{Conclusions and Future Work} \label{sec:conclusion}
The novel application of reinforcement learning to group structure optimization in neutron transport applications has shown to be effective and offers more flexibility than previously implemented optimization algorithms.
In this approach, an RL agent was trained to start from a high fidelity energy grid and remove group bounds until it reached a target number of groups. 
The starting number of groups varied when training the model, allowing for easily modifying the starting group structure and prevent retraining. 
At each step, the agent was given a reward based on the performance of the new energy grid and the number of energy groups. 
The error metric included the $k$-effective value and three reaction rates to ensure that the performance was not due to error cancellation in the flux spectra.
To accelerate the training and add stability , a 10 class classification surrogate model was introduced that took in the group structure as a vector of binary values and output the predicted class, with higher class numbers correlating to more accurate group structures. 
For non-homogeneous problems, the spatial widths and total cross sections are also included in the surrogate model as a separate branch of a multimodal neural network.  

This method was verified through two one-dimensional sphere problems, Godiva and the BeRP ball. 
For Godiva, the surrogate model was able accurately classify proposed grid structures 78\% of the time and an adjacent classification bin 98\% of the time. 
The BeRP ball surrogate model, which included spatial and neutron data, was able to achieve 70\% and 97\% true and adjacent accuracy. 
These provides an adequate method to evaluate energy grid structures without running full transport simulations.

Incorporating the surrogate models into the RL reward function in Eq.~\eqref{eq:r-class-2} accelerated the training process, resulting in less than 6 hours training on an Apple MacBook Pro versus over 8300 hours with the full transport simulations.
The trained RL models were verified using two different starting group structures, a 301 group and the LANL618, and two different targets, 30 and 70 group structures. 
The RL constructed group structures, RL30 and RL70, outperformed the LANL30 and LANL70 group structures in terms of both $k$-effective value and reaction rates when compared to a LANL618 group structure. 
The RL models also performed similarly to hierarchical agglomeration but requiring less computational time.

There are several directions to continue applying reinforcement learning to group structure optimization problems. 
Developing and improving on the surrogate models to incorporate material and spatial information for a variety of problems, materials, and problem layouts would help with the generalizability of this method. 
Restricting the classification bin sizes, especially for the higher accuracy models or implement residual RL will continue to improve the target group structure's performance. 
Expanding the abilities of the RL agents to not only remove bounds but to add or perturb energy bound locations will aid in the flexibility of this approach remove the restriction that group structures remain within the LANL618 group structure. 
The incorporation of multi-agent RL problems for group structure optimization should also be investigated.

\section*{Acknowledgments}
This work was supported by the U.S. Department of Energy through LANL. LANL is operated by Triad National Security, LLC, for the National Nuclear Security Administration of the U.S. Department of Energy (contract no. 89233218CNA000001).

\bibliographystyle{elsarticle-num} 
\bibliography{main}

@article{Berry:2021,
    title={Classification of group structures for a multigroup collision probability model using machine learning},
    author={Berry, Jessica J and Gil-Delgado, Gonzalo G and Osborne, Andrew GS},
    journal={Annals of Nuclear Energy},
    volume={160},
    pages={108367},
    year={2021},
    publisher={Elsevier}
}

@techreport{Berry:2022,
    title={Hierarchical Division and Clustering of Group Structures},
    author={Berry, Jessica Josephine and Saller, Thomas G},
    institution={Los Alamos National Laboratory},
    number={LA-UR-22-29528},
    year={2022},
}

@techreport{Bess:2019,
  title={The 2019 edition of the ICSBEP handbook},
  author={Bess, John D and Ivanova, Tatiana and Scott, Lori and Hill, Ian},
  year={2019},
  institution={Idaho National Laboratory (INL), Idaho Falls, ID (United States)}
}

@book{Chandrasekhar:1960,
    title={Radiative Transfer},
    author={Chandrasekhar, S.},
    year={1950},
    publisher={Dover},
    address={New York},
}

@inproceedings{Fasina:2022,
  address = {Pittsburgh, PA, USA},
  booktitle = {Proceedings of the PHYSOR-2022},
  title={Particle Swarm Optimisation for group structure optimization for radiotherapy shielding},
  author={Fasina, Oluwadamilola and Saller, Thomas},
  year={2022},
  publisher={American Nuclear Society}
}

@article{Gao:2024,
  title={Enhancing sample efficiency and exploration in reinforcement learning through the integration of diffusion models and proximal policy optimization},
  author={Gao, Tianci and Neusypin, Konstantin A and Dmitriev, Dmitry D and Yang, Bo and Rao, Shengren},
  journal={arXiv preprint arXiv:2409.01427},
  year={2024}
}

@article{Hochreiter:1997,
  title={Long short-term memory},
  author={Hochreiter, Sepp and Schmidhuber, J{\"u}rgen},
  journal={Neural computation},
  volume={9},
  number={8},
  pages={1735--1780},
  year={1997},
  publisher={MIT press}
}

@article{Huang:2020, 
    title={A Closer Look at Invalid Action Masking in Policy Gradient Algorithms}, 
    volume={35}, 
    DOI={10.32473/flairs.v35i.130584}, 
    journal={The International FLAIRS Conference Proceedings}, 
    author={Huang, Shengyi and Ontañón, Santiago}, 
    year={2022}, 
    month={May} 
}

@article{Johannink:2018,
  title={Residual reinforcement learning for robot control},
  author={Johannink, Tobias and Bahl, Shikhar and Nair, Ashvin and Luo, Jianlan and Kumar, Avinash and Loskyll, Matthias and Ojea, Juan Aparicio and Solowjow, Eugen and Levine, Sergey},
  journal={arXiv preprint arXiv:1812.03201},
  year={2018}
}

@book{Lewis:1993,
    Author = {E. E. Lewis and W. F. Miller},
    Publisher = {American Nuclear Society Scientific Publications},
    Title = {Computational Methods of Neutron Transport},
    Year = {1993}
}

@ARTICLE{Li:2022,
  author={Li, Zewen and Liu, Fan and Yang, Wenjie and Peng, Shouheng and Zhou, Jun},
  journal={IEEE Transactions on Neural Networks and Learning Systems}, 
  title={A Survey of Convolutional Neural Networks: Analysis, Applications, and Prospects}, 
  year={2022},
  volume={33},
  number={12},
  pages={6999-7019},
}

@techreport{MacFarlane:1987,
  title={The NJOY Nuclear Data Processing System: Volume 3, The GROUPR, GAMINR, and MODER modules},
  author={MacFarlane, Robert E and Muir, Douglas W},
  year={1987},
  institution={Los Alamos National Lab.(LANL), Los Alamos, NM (United States)}
}

@techreport{Njoy:2017,
    title={The {NJOY} nuclear data processing system, version 2016},
    author={Macfarlane, R and Muir, D W and Boicourt, R M and Kahler III, A C and Conlin, J L},
    institution={Los Alamos National Laboratory}, 
    number = {LA-UR-17-20093},
    year={2017},
}

@techreport{Rouse:2025,
    author={Rouse, Natalie Kealaula and Whewell, Benjamin Joseph and Gibson, Nathan Andrew},
    title={MetaHeuristic Feature Selection for Energy Group Optimization and Analysis},
    institution={Los Alamos National Laboratory},
    number={LA-UR--25-29285},
    year={2025}
}

@article{Saller:2023,
  title={Using a random forest model to choose optimized group structures},
  author={Saller, Thomas G and Nair, Vishnu and Till, Andrew and Gibson, Nathan},
  journal={Nuclear science and engineering},
  volume={197},
  number={8},
  pages={2117--2135},
  year={2023},
  publisher={Taylor \& Francis}
}

@article{Schulman:2017,
    title={Proximal Policy Optimization Algorithms},
    author={Schulman, John and Wolski, Filip and Dhariwal, Prafulla and Radford, Alec and Klimov, Oleg},
    journal={arXiv preprint arXiv:1707.06347},
    year={2017}
}

@article{stable-baselines3,
    author  = {Antonin Raffin and Ashley Hill and Adam Gleave and Anssi Kanervisto and Maximilian Ernestus and Noah Dormann},
    title   = {Stable-Baselines3: Reliable Reinforcement Learning Implementations},
    journal = {Journal of Machine Learning Research},
    year    = {2021},
    volume  = {22},
    number  = {268},
    pages   = {1-8},
    url     = {http://jmlr.org/papers/v22/20-1364.html}
}

@book{Sutton:2020,
    title={Reinforcement Learning: An Introduction},
    author={Sutton, Richard S and Barto, Andrew G},
    edition={Second},
    year={2020},
    publisher={The MIT Press},
    address={Cambridge, Massachusetts, USA}
}

@article{Wimett:1960,
    author = {T. F. Wimett and R. H. White and W. R. Stratton and D. P. Wood},
    title = {Godiva II—An Unmoderated Pulse-Irradiation Reactor*},
    journal = {Nuclear Science and Engineering},
    volume = {8},
    number = {6},
    pages = {691--708},
    year = {1960},
    publisher = {Taylor \& Francis},
}

@article{Yi:2013,
  title={Energy group structure determination using particle swarm optimization},
  author={Yi, Ce and Sjoden, Glenn},
  journal={Annals of Nuclear Energy},
  volume={56},
  pages={53--56},
  year={2013},
  publisher={Elsevier}
}

\end{document}